\documentclass[journal]{IEEEtran}
\ifCLASSINFOpdf
\else
\fi
\usepackage{amsmath,amssymb}
\usepackage{bm}
\usepackage{graphicx}
\usepackage{subfigure}
\usepackage{float}
\usepackage{subfig}
\usepackage{booktabs}
\begin{document}
%

\title{Hybrid Beamforming with Orthogonal delay-Doppler Division Multiplexing Modulation for Terahertz Sensing and Communication}

%
%
%

\author{Meilin~Li,~\IEEEmembership{Graduate Student Member,~IEEE,}
        Chong~Han,~\IEEEmembership{Senior Member,~IEEE,} Shi~Jin,~\IEEEmembership{Fellow,~IEEE}
        
\thanks{This paper was presented in part at IEEE GLOBECOM, December 2024~\cite{li_offgrid_2024}.

Meilin Li and Chong Han are with Terahertz Wireless Communications (TWC) Laboratory, Shanghai Jiao Tong University, Shanghai 200240. E-mail: \{meilinli, chong.han\}@sjtu.edu.cn.}
\thanks{Shi Jin is with the National Mobile Communications Research Laboratory, Southeast University, Nanjing 210096, China. E-mail: jinshi@seu.edu.cn.}
}

\maketitle

\begin{abstract}
The Terahertz band holds a promise to enable both super-accurate sensing and ultra-fast communication. However, challenges arise that severe Doppler effects call for a waveform with high Doppler robustness while severe propagation path loss urges for an ultra-massive multiple-input multiple-output (UM-MIMO) structure. To tackle these challenges, hybrid beamforming with orthogonal delay-Doppler multiplexing modulation (ODDM) is investigated in this paper. 
First, the integration of delay-Doppler waveform and MIMO is explored by deriving a hybrid beamforming-based UM-MIMO ODDM input-output relation.
Then, a multi-dimension sensing algorithm on target azimuth angle, elevation angle, range and velocity is proposed, which features low complexity and high accuracy. 
Finally, a sensing-centric hybrid beamforming is proposed to design the sensing combiner by minimizing the Cram\'er-Rao lower bounds (CRLB) of angles. After that, the precoder that affects both communication and sensing is then designed to maximize the spectral efficiency. Numerical results show that the sensing accuracy of the proposed sensing algorithm is sufficiently close to CRLB. Moreover, the proposed hybrid beamforming design allows to achieve maximal spectral efficiency, millimeter-level range estimation accuracy, millidegree-level angle estimation accuracy and millimeter-per-second-level velocity estimation accuracy.
Take-away lessons are two-fold. Combiner design is critical especially for sensing, which is commonly neglected in hybrid beamforming design for communication. 
Furthermore, the optimization problems for communication and sensing can be decoupled and solved independently, significantly reducing the computational complexity of the THz monostatic ISAC system.
\end{abstract}

\begin{IEEEkeywords}
Terahertz integrated sensing and communication, Orthogonal delay-Doppler multiplexing modulation, hybrid beamforming, Cram\'er-Rao lower bounds.
\end{IEEEkeywords}

%
\IEEEpeerreviewmaketitle

\section{Introduction}
\subsection{Motivation}
 \IEEEPARstart{T}{erahertz} (THz) band (0.1-10 THz)~\cite{akyildiz2014terahertz} was long considered as the THz gap and remained largely unexplored for many years, due to the lack of efficient sources and detectors. Recent breakthroughs have bridged this gap, unlocking ultra-high-capacity wireless backhaul links~\cite{sen2023multi,liu2024high}. The concept of Terahertz Integrated Sensing and Communication (THz ISAC)~\cite{han_thz_2024} has emerged as a promising dual-purpose solution. As a high-frequency spectrum, the THz band offers significant potential to facilitate both sensing and communication~\cite{liu2022integrated}. With a tenfold increase in bandwidth and antenna arrays up to ten times larger than those used in millimeter-wave communications~\cite{chen2019survey}, the THz band can support wireless links with data rates reaching hundreds of gigabits per second. Additionally, the broader bandwidth and larger antenna arrays enable enhanced range and angular resolution, enabling millimeter-level sensing accuracy.

Despite advancements, several concerns persist due to the nature of Terahertz band, such as severe Doppler effects. Both the higher carrier frequency in the Terahertz band and mobile scenarios in future wireless network~\cite{jornet2024mobile,buzzi2023leo} can induce severer Doppler effects. Orthogonal frequency division multiplexing (OFDM) modulation is a widely-used technique, renowned for using cyclic prefix to combat the effects of frequency-selective channels, which are typically caused by multipath propagation and delay spread. However, OFDM fails to address the effects of doubly-selective channels, which are additionally caused by Doppler spread. If OFDM is regularly deployed in Terahertz communication, the severe Doppler spread will cause frequency shifts among subcarriers, which breaks orthogonality and brings inter-carrier interference (ICI)~\cite{wu_dft-spread_2023}. Thus, the first issue to be addressed is the Doppler robustness, both in communication and sensing cases.

A second major challenge in the Terahertz band is severe path loss. Terahertz band offers substantial bandwidth at the cost of limited transmission distance and high propagation loss. However, the short wavelengths at these frequencies facilitate the deployment of ultra-massive MIMO (UM-MIMO) antenna arrays, which can generate sharp beams with high beamforming gains~\cite{chen_millidegree-level_2022}. Hybrid beamforming presents an efficient alternative to fully digital beamforming by reducing the number of required radio frequency (RF) chains for managing and coordinating individual antennas. Given that UM-MIMO systems can deploy up to thousands of antennas, hybrid beamforming-based design with low computational complexity is essential for realizing the potential of THz-based wireless systems.

The third challenge lies in the open question on how to unify and realize integrated sensing and communication. From coexistence to full-fledged integration, sensing and communication can achieve more integration gain with more sufficient exploitation of the limited hardware~\cite{lu2024integrated}. Among others, a well-designed waveform for both functionalities, which is supposed to have great compatibility with MIMO, as a higher level of integration, promises to be an optimal solution for both functionalities. In terms of THz ISAC, there is a strong urge to find a waveform with high Doppler robustness and compatible with hybrid beamforming-based UM-MIMO architecture.

\subsection{Related Works}
\subsubsection{Waveform design}
As for the waveform design for THz communication, high spectral efficiency, low peak-to-average power ratio (PAPR) as well as high Doppler robustness are three main considerations. When it comes to THz ISAC waveforms, \cite{han_thz_2024} lists four main candidates, including two time-frequency domain waveforms, OFDM and DFT spread OFDM, as well as two delay-Doppler domain waveforms, orthogonal time frequency space (OTFS)~\cite{hadani_orthogonal_2017} and DFT spread OTFS~\cite{wu_dft-spread_2023}. The DFT-spread procedure significantly reduces PAPR, thereby enhancing signal quality after passing through amplifiers. In contrast to time-frequency domain waveforms, delay-Doppler domain waveforms exploit the property of doubly selective channel in the delay-Doppler domain, providing a strong delay-resilience and Doppler-resilience. In this case, delay-Doppler modulation shows great robustness towards Doppler effects. Orthogonal delay-Doppler division multiplexing (ODDM) modulation is proposed as a practical delay-Doppler modulation by applying square root raised cosine pulse~\cite{tong_orthogonal_2024}, which is proved to be a delay-Doppler domain orthogonal pulse in \cite{lin_orthogonal_2022}, and exhibits lower out-of-band emission compared with OTFS. 

Numerous studies explore how delay-Doppler modulation can support both functionalities of communication and sensing.  A significant portion of this research focuses on low-complexity signal detection problem, by proposing a message passing algorithm~\cite{raviteja_interference_2018}, a MP-based soft-output detector~\cite{gaudio_effectiveness_2020}, a linear equalizer~\cite{surabhi_low-complexity_2020}, a conjugate gradient based equalizer~\cite{wu_dft-spread_2023}, or by contrastive learning~\cite{cheng_novel_2023}, etc. Other studies consider to perform sensing paramater estimation, in channel estimation or in monostatic sensing, by developing a maximum likelihood estimator~\cite{gaudio_effectiveness_2020,wu_dft-spread_2023,wang_exploring_2024} or compressed sensing~\cite{gomez-cuba_compressed_2021}.

Compared to research in single-input single-output (SISO) systems, significantly fewer studies focus on the compatibility of delay-Doppler modulation with multiple-input multiple-output (MIMO) systems. ODDM-based signal detection based on generative adversarial network~\cite{cheng_mimo-oddm_2024} and OTFS-based MIMO radar estimation~\cite{dehkordi_beam-space_2023} have been investigated. \cite{li_novel_2022} proposes a spatially spread OTFS approach for angular domain discretization, aiming to simplify receive signal processing. These studies highlight the growing interest in integrating delay-Doppler modulation with MIMO systems in the context of sensing and communication, yet much remains to be explored in this area.

\subsubsection{Beamforming design}
The beamforming design for sensing completely differs from the beamforming design for communication in the following three aspects.
Firstly, it is inappropriate to assume known angle information of the target, since the task of sensing implies that none of the information of the target is known to the transceiver. However, most beamforming optimization problems assume target angle is known. In most studies, they assume approximate angle information~\cite{song_intelligent_2023,liu_cramer-rao_2022} in target tracking condition or assume an omni-directional search before beamforming design~\cite{elbir_terahertz-band_2021}. 

Secondly, the optimization metrics for sensing are different from those for communication. Widely used metrics for sensing include the mean square error (MSE) between the ideal and designed beamforming matrix~\cite{wu_time-frequency-space_2024,elbir_terahertz-band_2021,yu_hybrid_2023} or beam pattern~\cite{qi_hybrid_2022}, as well as sensing signal-to-noise ratio (SNR)~\cite{qian_sensing-based_2023}. These metrics are based on the fundamental assumption that sensing performance improves by maximizing signal power in the target direction\cite{luo_joint_2022}. \cite{liu_cramer-rao_2022} theoretically validates this idea by minimizing the Cram\'er Rao Lower Bound (CRLB) subjecting to constraints on communication SINR. Then, CRLB has become a novel metric for sensing performance optimization~\cite{song_intelligent_2023,hua2023mimo}.

Lastly, most studies focus on the beamforming design of the precoder, or in other words, transmit design~\cite{wu_time-frequency-space_2024,babu2024precoding}. However, while communication focuses on recovering signals in the presence of noise, sensing aims to achieve precise estimation under noise interference. Comparatively, sensing relies more heavily on the accurate modeling of noise covariance. Thus, combiner design for sensing is much more important than that for communication. 

\subsection{Contributions}
The contributions of this work can be summarized as follows:
\begin{itemize}
    \item \textbf{We develop the integration of MIMO and delay-Doppler waveforms and derive a hybrid beamforming-based UM-MIMO ODDM input-output relation.} First, an off-grid single-input single-output relation is derived to break the sensing resolution limitations brought by on-grid delay and Doppler assumption in most existing studies. Then, we consider to apply hybrid beamforming structure, instead of analog or digital beamforming, for the sake of generality. We propose a UM-MIMO ODDM hybrid beamforming structure and derive its input-output relation.
    \item \textbf{We derive the CRLB of azimuth angle, elevation angle, range and velocity of the target and propose a multi-dimension sensing algorithm that can approach CRLBs.} To realize the functionality of sensing, we propose a sensing algorithm to estimate the multi-dimension parameters, including the azimuth angle, elevation angle, range and velocity of the target, by deriving a maximum likelihood estimator. The proposed sensing algorithm reduces computational complexity by exploiting the property of circular matrix and reduces sample complexity by designing efficient search strategy over four dimensional variables. Besides, we derive the CRLB for estimating four target parameters in hybrid beamforming structure. Numerical results show that the sensing accuracy of proposed algorithm is sufficiently close to its theoretical limits, demonstrating the high accuracy of the proposed algorithm.
    \item \textbf{We propose an optimal solution of hybrid beamforming design for THz ISAC.} 
    First, we derive the CRLB of azimuth angle and elevation angle as a novel metric for optimizing sensing performance. 
    Building upon these analytical derivations, we propose a sensing-centric hybrid beamforming design that optimizes the sensing combiner, and propose a corresponding sensing combiner optimization algorithm. This algorithm is able to design the combiner matrix based on arbitrary angles and regenerate to sweep the angular area. Furthermore, based on the optimized sensing combiner, the precoder is further designed to maximize spectral efficiency. Numerical results show that by designing the precoder and communication combiner to maximize spectral efficiency, and optimizing sensing combiner to minimize CRLB, we can achieve maximal spectral efficiency and maintain millimeter-level sensing accuracy.
\end{itemize}

The structure of this paper is organized as follows. The system framework with signal model derivation of hybrid beamforming-based UM-MIMO ODDM is presented in Sec.~\ref{sec:system}. The sensing algorithms and the hybrid beamforming design are proposed in Sec.~\ref{sec:sensing_algo} and Sec.~\ref{sec:beamforming_design}, respectively. Sec.~\ref{sec:simulation} illustrates performance evaluation. Finally, the paper is concluded in Sec.~\ref{sec:conclusion}.

\section{System Model}\label{sec:system}
\begin{figure}[t]
    \centering
    \includegraphics[width=0.8\linewidth]{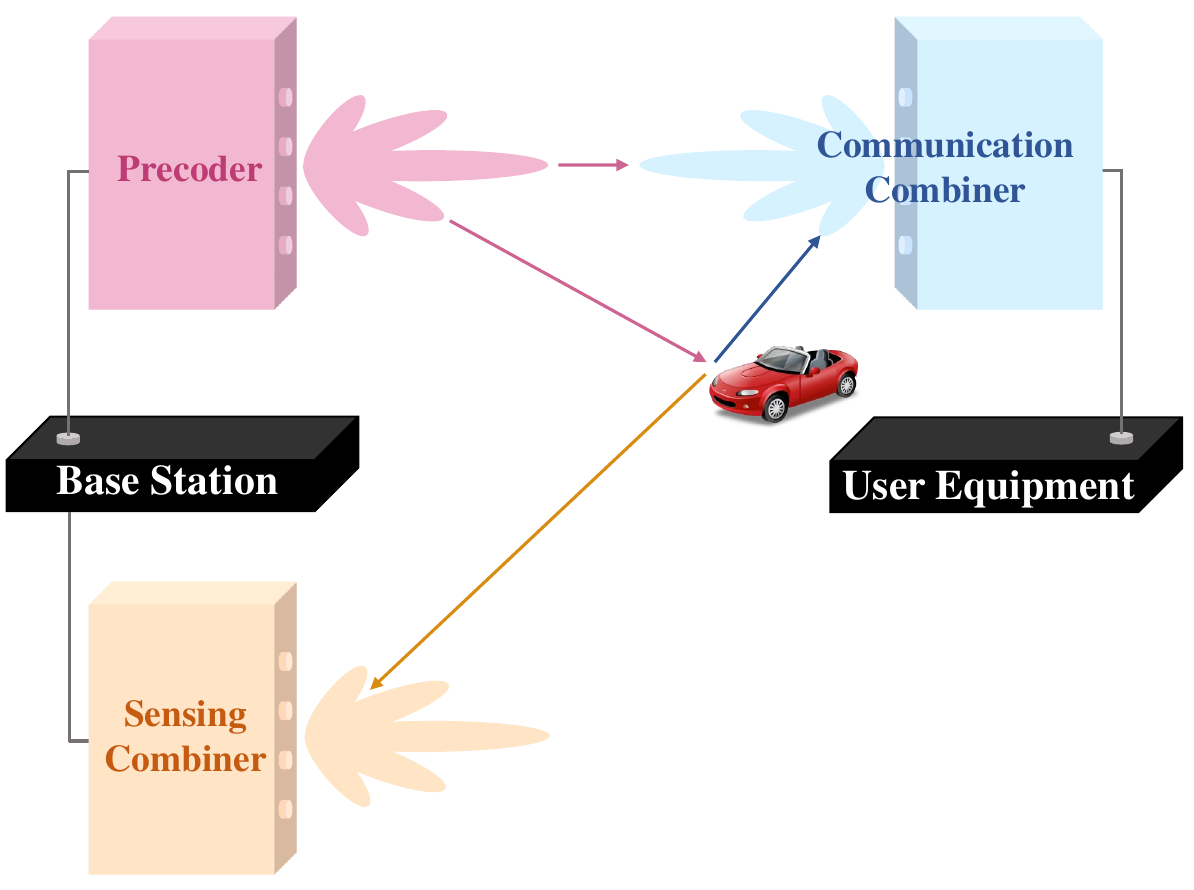}
    \caption{Concept diagram for Terahertz monostatic sensing and communication.}
    \label{fig: concept diagram}
    \vspace{-5mm}
\end{figure}
In this section, we consider a THz UM MIMO ODDM monostatic sensing and communication, as depicted in Fig. \ref{fig: concept diagram}. The base station is equipped with co-located transmitter (TX) and receiver (RX), to transmit the signal and process the back-scattered signal from the target. 
The TX-to-RX leakage can be mitigated by spatial separation or self-interference cancellation~\cite{masmoudi2016channel} and is ignored here for simplicity.
We will begin with the derivation of single-input single-output relation which allows off-grid delay and Doppler shifts. Afterwards, it will be extended to a fully-connected hybrid beamforming structure of the UM MIMO system.

\subsection{Off-grid Single-Input Single-Output Relation}\label{sec2A}

\begin{figure*}
    \centering
    \includegraphics[width=0.8\linewidth]{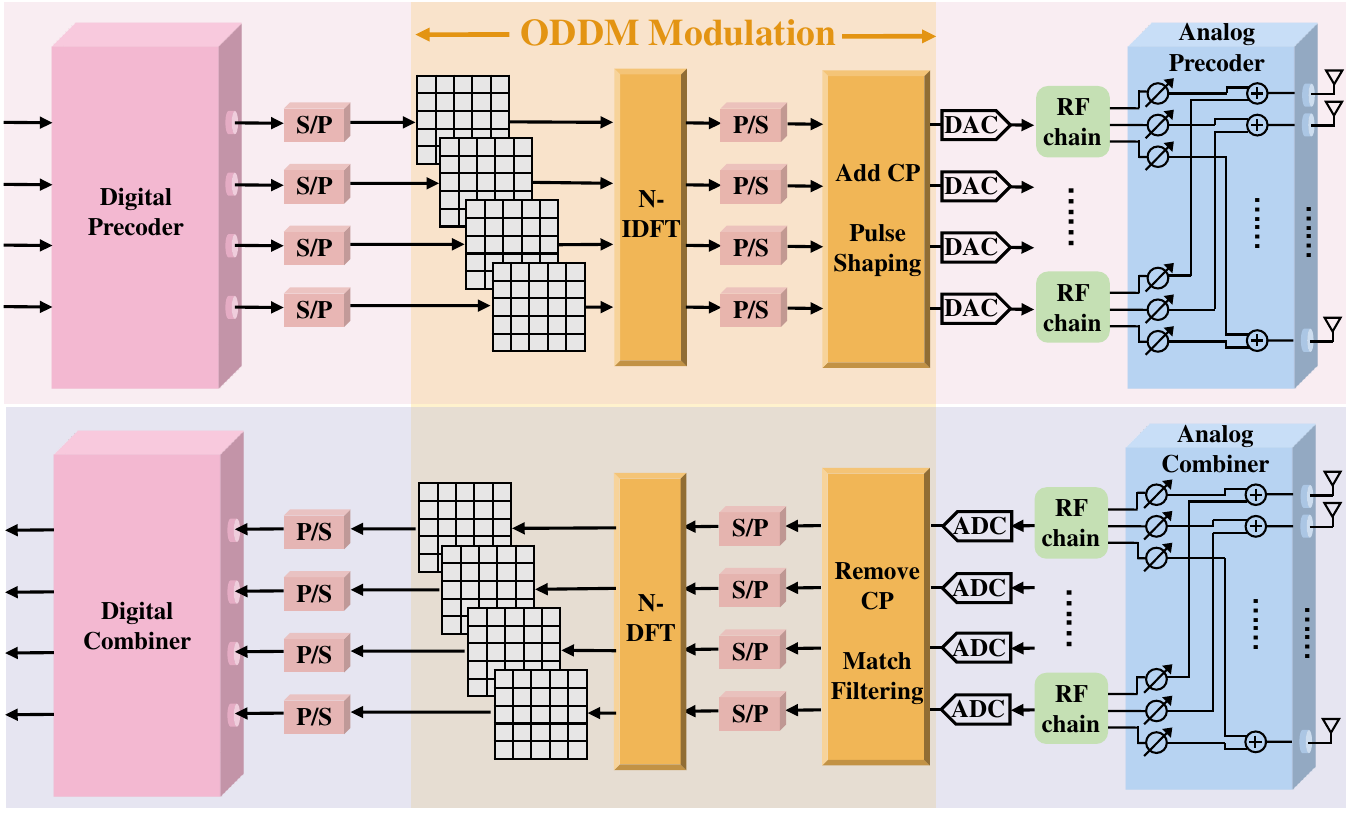}
    \caption{UM MIMO ODDM hybrid beamforming structure.}
    \label{fig: system model}
    \vspace{-5mm}
\end{figure*}

Information symbols are arranged in an ODDM frame with $M$ delay bins and $N$ Doppler bins. The resolutions of delay and Doppler shifts are $\frac{T}{M}$ and $\frac{1}{NT}$, where $T = \frac{1}{\Delta f}$ and $\Delta f$ is the subcarrier spacing. Let $\mathbf{X}^{\text {DD}} \in\mathbb{C}^{M\times N} $ denote the $M \times N$ information symbols in the delay-Doppler domain. $\mathbf{X}^{\text {DD}}[m, n]$ denotes the symbol in the ${m}^\text{th}$ delay bin and the ${n}^\text{th}$ Doppler bin ($0 \leq m \leq M-1$, $0 \leq n \leq N-1$). 

Next, the transmitter transforms the delay-Doppler domain symbols to delay-time domain symbol by $N$-point IDFT and staggers the samples at an interval of $\frac{T}{M}$. The time domain transmitted $\mathbf {x}^{\text {TD}}$ is given by
\begin{equation} 
	\mathbf {x}^{\text {TD}} = \text {vec}(\mathbf {X}^{\text {DD}} \mathbf {F}_{N}^{H})= \left({\mathbf {F}_{N}^{H} \otimes \mathbf {I}_{M} }\right) \text {vec} (\mathbf {X}^{\text {DD}}), 
\end{equation}
where $\mathbf {F}_{N}\in \mathbb{C}^{N\times N}$ describes the normalized DFT matrix.

We consider a square root raised cosine pulse $a(t)$, which is proved to be a delay-Doppler plane orthogonal pulse~\cite{lin_orthogonal_2022}. In this case, $a(t)$ spans a time duration of $2Q\frac{T}{M}$ ($Q < \frac{M}{2}$). After pulse shaping, the baseband time-domain transmitted signal can be expressed as
\begin{equation} 
	x(t) =\sum_{q=0}^{M N-1} x[q] a\left(t-q \frac{T}{M}\right). 
\end{equation}
where $x[q]$ is the $q^\text{th}$ element of the time domain transmitted vector $\mathbf {x}^{\text {TD}}$ ($0 \leq q \leq M N-1$). Then, one cyclic prefix is inserted into time-domain transmitted signal for each ODDM frame, as
\begin{equation} 
	x_{cp}(t) =\sum_{q=-M_{\text{cp}}}^{M N-1} x[q] a\left(t-q \frac{T}{M}\right), 
\end{equation}

The sensing channel is considered as a doubly selective channel including $P$ targets, whose impulse response is given by 
\begin{equation} h(\tau, \nu) = \sum _{p=1}^{P} \alpha _{p} \delta (\tau - \tau _{p}) \delta (\nu - \nu _{p}), 
\end{equation}
where $\alpha_p$ denotes the complex path coefficient of ${p}^\text{th}$ target and is subject to the power attenuation characteristics of the THz band~\cite{han2014multi}. The range and velocity of the ${p}^\text{th}$ target relative to sensing receiver are denoted by $r_p$ and $v_p$ ($1 \leq p \leq P$). Thus, the delay and Doppler shift of the echo signal can be calculated as $\tau_p = \frac{2r_p}{c_0}$, $\nu_p = \frac{2 f_c v_p}{c_0}$($\tau_p \in \left.(0,M_{\text{cp}}\frac{T}{M}\right.]$, $ \nu_p \in \left(-\frac{1}{2T},\frac{1}{2T}\right]$), where $c_0$ represents light speed and $f_c$ represents the carrier frequency. 

The matched filter at the receiver side is $a^{\ast}(-t)$, so that the output of matched filtering is the convolution between received signal $r(t)$ and matched filter $a^{\ast}(-t)$, given by
\begin{equation} y(t) = r(t) \ast a^{\ast}(-t) = h(\tau, \nu) \ast x_{cp}(t) \ast a^{\ast}(-t) +z(t),
	\label{equ5}
\end{equation}
where $z(t)$ represents the noise signal, and $x_{cp}(t)$ is given by
\begin{equation}
	x_{cp}(t) = \left(\sum_{q=-M_{\text{cp}}}^{M N-1} x[q] \delta\left(t-q \frac{T}{M}\right)\right) \ast a(t).
	\label{equ6}
\end{equation}
By substitution of (\ref{equ6}), (\ref{equ5}) can be rewritten as
\begin{align} 	
	\!	y(t) \! =  \! \sum_{q=-M_{\text{cp}}}^{M \! N-1} \! \sum_{p=1}^P \! \alpha_p x\!\left[q\right] g\left(\! t-q  \frac{T}{M}-\tau_p\!\right) e^{j 2 \pi \nu_p t}\!+\!z(t),
\end{align}
where $g(t)\triangleq a(t)*a^*(-t)$~\cite{tong_orthogonal_2024}.

The received signal $ y(t)$ is sampled at $t = k T_s$, where $T_s = \frac{T}{M}$ denotes sampling period. Thus the received samples can be given by
\begin{equation}
    \! y[k] \! = \! \sum_{q\!=-M_{\!\text{cp}}}^{M \! N-1} \! \sum_{p=1}^P \alpha_p x \!\left[q\right]\!  g \!\left(k T_s\! - q T_s\! -\tau_p\!\right) \! e^{j 2 \pi \nu_p k T_s} \! + \! z[k]\!.
	\label{equ8}
\end{equation} 	

Let $l_p = \frac{\tau_p}{T_s}$ and $k_p = \nu_p M N T_s$ denote the normalized delay and Doppler shift of the ${p}^\text{th}$ path respectively. Note that we make a more general assumption that $l_p$ and $k_p$ are not necessary to be integers. Thus, (\ref{equ8}) can be rewritten as
\begin{equation}
	y[k] \! = \! \sum_{q=-M_{\mathrm{cp}}}^{M\!N-\!1} \! \sum_{p=1}^P \! \alpha_px[q]g\left(\left(k\!-q\!-l_p\right)T_s\right)e^{j2\pi\frac{k_p k}{MN}} \! + \! z[k] \!.
	\label{equ9}	
\end{equation}
Finally, (\ref{equ9}) can be evolved in matrix format, as
\begin{equation}
	\mathbf y^{\text {TD}} = \mathbf H^{\text {TD}}_{{\text {s}}} \mathbf {x}^{\text {TD}}+ \mathbf z=  \sum_{p=1}^P\alpha_p \mathbf{\Delta}_p \mathbf{G}_{l_p} \mathbf {x}^{\text {TD}}+ \mathbf z,
	\label{equ10}
\end{equation}
where $\mathbf y^{\text {TD}}={y[k]}_{0 \leq k \leq MN-1}$ denote the time domain receive samples, and $\mathbf H^{\text {TD}}_{{\text {s}}}$ represents the single-input single-output time domain channel. $\mathbf{\Delta}_p$ is given by
\begin{equation}
	{{\mathbf{\Delta}_p }} = \text{diag}\left[ z^{0}, z^{1}, \ldots, z^{MN-1} \right],
\end{equation}
with $z = e^{j\frac{2\pi}{MN} k_p}$. Since path delay $\tau_p \in \left(0,M_{\text{cp}}\frac{T}{M}\right]$, the normalized delay satisfies $l_p < M_{\text{cp}}$. In this case, $\mathbf{G}_{l_p}$ is a circular matrix, in which the element of $\mathbf{G}_{l_p}$ can be written as 
\begin{equation}
	\mathbf{G}_{l_p}[k,q] = g \left(\left( \left[k-q-l_p+M_{\text{cp}}\right] _{MN}-M_{\text{cp}}\right) T_s\right).
\end{equation} 
Note that $g(t)$ is the raised cosine function which spans a time duration of $2Q\frac{T}{M}$. Thus, only $(2Q+1)$ out of $MN$ elements are nonzero in each row of matrix $\mathbf{G}_{l_p}$. Here, $\mathbf{G}_{l_p}$ and $\mathbf{\Delta}_p$ model the effect of delay and Doppler shift, respectively.

Particularly, if $l_p$ and $k_p$ are integers, which means the path delay and Doppler shift are integer multiples of their corresponding resolution, the circular matrix $\mathbf{G}_{l_p}$ evolves into ${{\mathbf {\Pi }}}^{l_{p}}$, and (\ref{equ10}) turns into
\begin{equation}
	\mathbf y^{\text {TD}} = \sum_{p=1}^P \alpha_p {{\bf \Delta }}^{k_{p}} {{\bf {\Pi }}}^{l_{p}} \mathbf {x}^{\text {TD}}+ \mathbf z, 
	\label{equ14}
\end{equation}
where ${{\bf \Delta }} = \text{diag}\left[ z^{0}, z^{1}, \ldots, z^{MN-1} \right]$ and $z = e^{j\frac{2\pi}{MN}}$. (\ref{equ14}) is consistent with the case of OTFS.

Time domain receive vector can be rearranged in matrix format $\mathbf {Y}^{\text {TD}} =\text {vec}^{-1}(\mathbf {y}^{\text {TD}})$. It can be transformed into delay-Doppler domain via $N$-point DFT, as
\begin{equation}
	\mathbf {y}^{\text {DD} }= \text {vec}(\mathbf {Y}^{\text {DD}}) =\left({\mathbf {F}_{N} \otimes \mathbf {I}_{M} }\right) \mathbf {y}^{\text {TD}}. 
\end{equation}
where $\mathbf {Y}^{\text {DD}} \in \mathbb{C}^{M\times N}$ and $\mathbf {y}^{\text {DD} }\in \mathbb{C}^{MN \times 1}$ denote the delay-Doppler domain received signal matrix and vector.

Finally, the delay-Doppler domain input/output relation can be expressed as 
\begin{equation}
	\mathbf {y}^{\text {DD} }= \sum_{p=1}^P \alpha_p \left({\mathbf {F}_{N} \otimes \mathbf {I}_{M} }\right) \mathbf{\Delta}_p \mathbf{G}_{l_p} \left( {\mathbf {F}_{N}^{H} \otimes \mathbf {I}_{M} }\right) \mathbf {x}^{\text {DD}} + 
	\tilde{\mathbf z},
\end{equation}
where $\tilde{\mathbf z} = \left({\mathbf {F}_{N} \otimes \mathbf {I}_{M} }\right) \mathbf z$.

\begin{figure*}[!t] 
	\normalsize
	\begin{align*}
	\mathbf {y}^{\text {DD}} 
	&= \left(
	\left( \mathbf{W}_{\text{RF}} \mathbf{W}_{\text{BB}}\right)^{H}  \otimes 
	\left( {\mathbf {F}_{N} \otimes \mathbf {I}_{M} } \right) 
	\right)
	\mathbf H^{\text {TD}}_{{\text {m}}}
	\left( \mathbf{F}_{\text{RF}} \mathbf{F}_{\text{BB}} \otimes 
	\left( {\mathbf {F}_{N}^{H} \otimes \mathbf {I}_{M} } \right) 
	\right)
	\mathbf {x}^{\text {DD}} + \mathbf {n}^{\text {DD}} \\
	&= \sqrt {{N_{t}}{N_{r}}} \sum_{p=1}^P \alpha_p \left(
	\left( \mathbf{W}_{\text{RF}} \mathbf{W}_{\text{BB}}\right)^{H}  \otimes 
	\left( {\mathbf {F}_{N} \otimes \mathbf {I}_{M} } \right) 
	\right)
	\left( \mathbf {A}(\theta_p,\phi_p) \otimes \mathbf{\Delta}_p \mathbf{G}_{l_p}\right) 
	\left( \mathbf{F}_{\text{RF}} \mathbf{F}_{\text{BB}} \otimes 
	\left( {\mathbf {F}_{N}^{H} \otimes \mathbf {I}_{M} } \right) 
	\right)
	\mathbf {x}^{\text {DD}} + \mathbf {n}^{\text {DD}} \\
	&= \sqrt {{N_{t}}{N_{r}}} \sum_{p=1}^P \alpha_p \left(
	\left( \mathbf{W}_{\text{RF}} \mathbf{W}_{\text{BB}}\right)^{H} \mathbf {A}(\theta_p,\phi_p)
	\mathbf{F}_{\text{RF}} \mathbf{F}_{\text{BB}}
	\right)
	\otimes 
	\left( \left( {\mathbf {F}_{N} \otimes \mathbf {I}_{M} }\right)  \mathbf{\Delta}_p \mathbf{G}_{l_p}
	\left( {\mathbf {F}_{N}^{H} \otimes \mathbf {I}_{M} }\right) 
	\right) 
	\mathbf {x}^{\text {DD}} + \mathbf {n}^{\text {DD}},
	\tag{28}
	\label{equ29}
	\end{align*}
    \hrulefill
\end{figure*}

\subsection{Multiple-Input Multiple-Output Monostatic Radar Model}
We consider a fully-connected hybrid beamforming structure of UM-MIMO system shown in Fig.~\ref{fig: system model}. The system generates $N_s$ data streams, denoted as $\mathbf {X}^{\text {DD}} \in \mathbb{C}^{MN \times N_s}$. At the transmitter side, the delay-Doppler domain data stream is precoded through a digital precoder $\mathbf{F}_{\text{BB}} \in \mathbb{C}^{N_\text{RF} \times N_s}$, where $N_\text{RF}$ refers to the number of RF chains. Let $\tilde{\mathbf {X}}^{\text {DD}}$ denote the precoded delay-Doppler data streams, given by
\begin{equation}
	\tilde{\mathbf {X}}^{\text {DD}} = \mathbf{F}_{\text{BB}} \left[ \mathbf {X}^{\text {DD}} \right]^{T} \in \mathbb{C}^{N_\text{RF} \times MN},
\end{equation}
Then, we perform ODDM modulation to transform the delay-Doppler domain data blocks into time domain samples. Let $\tilde{\mathbf {X}}_{i}^{\text {DD}} \in \mathbb{C}^{M \times N}$ and $\mathbf {\tilde X}_{i}^{\text {TD}} \in \mathbb{C}^{M \times N}$ denote the delay-Doppler and time domain matrix for ${i}^\text{th}$ data stream ($1 \leq i \leq N_s$). Thus, their relation can be written as
\begin{equation}
	\mathbf {\tilde {X}}_{i}^{\text {TD}} = \tilde{\mathbf {X}}_{i}^{\text {DD}} \mathbf {F}_{N}^{H},
	\label{equ18}
\end{equation}
We can vectorize (\ref{equ18}) as
\begin{equation}
	\mathbf {\tilde x}_{i}^{\text {TD}} = \left({\mathbf {F}_{N}^{H} \otimes \mathbf {I}_{M} }\right) \tilde{\mathbf {x}}_{i}^{\text {DD}},
	\label{equ19}
\end{equation}
Let $\mathbf {\tilde X}^{\text {TD}} = [\mathbf {\tilde x}_{1}^{\text {TD}},\cdots,\mathbf {\tilde x}_{N_s}^{\text {TD}}]$ denote the time-domain discrete signal after modulated with ODDM. Then $\mathbf {X}^{\text {TD}}$ can be calculated as
\begin{align*}
	\mathbf {\tilde X}^{\text {TD}} 
	& = \left({\mathbf {F}_{N}^{H} \otimes \mathbf {I}_{M} }\right) [\mathbf {\tilde x}_{1}^{\text {DD}},\cdots,\mathbf {\tilde x}_{N_s}^{\text {DD}}] \\
	& = \left({\mathbf {F}_{N}^{H} \otimes \mathbf {I}_{M} }\right) \left[ \mathbf {\tilde X}^{\text {DD}} \right]^{T} \\
	& = \left({\mathbf {F}_{N}^{H} \otimes \mathbf {I}_{M} }\right) \mathbf {X}^{\text {DD}} \left[ \mathbf{F}_{\text{BB}}\right]^{T},\tag{19}
    \setcounter{equation}{19}
\end{align*}

An analog precoder $\mathbf{F}_{\text{RF}} \in \mathbb{C}^{N_t \times N_\text{RF}}$ is placed after conducting ODDM modulation, where $N_t$ represents the number of transmit antennas. The transmit time-domain signal, whose matrix format $\mathbf {X}^{\text {TD}} \in \mathbb{C}^{MN \times N_t}$ and vector format $\mathbf {x}^{\text {TD}} \in \mathbb{C}^{MN N_t\times 1}$, can be respectively expressed as
\begin{equation}
	\mathbf {X}^{\text {TD}} = \left({\mathbf {F}_{N}^{H} \otimes \mathbf {I}_{M} }\right) \mathbf {X}^{\text {DD}} \left[ \mathbf{F}_{\text{BB}}\right]^{T} \left[ \mathbf{F}_{\text{RF}}\right]^{T},
\end{equation}
and
\begin{equation}
	\mathbf {x}^{\text {TD}} = \left( \mathbf{F}_{\text{RF}} \mathbf{F}_{\text{BB}} \otimes 
	\left( {\mathbf {F}_{N}^{H} \otimes \mathbf {I}_{M} } \right) 
	\right) \mathbf {x}^{\text {DD}}.
	\label{equ22}
\end{equation}

Equivalently, the co-located sensing receiver is equipped with an analog combiner $\mathbf{W}_{\text{RF}} \in \mathbb{C}^{N_r \times N_\text{RF}}$ and a digital combiner $\mathbf{W}_{\text{BB}} \in \mathbb{C}^{N_\text{RF} \times N_s}$. Similar with the transmitter, the received time-domain vector and the combined delay-Doppler domain vector are denoted as $\mathbf {y}^{\text {TD}}$ and $\mathbf {y}^{\text {DD}}$. Their relationship can be written as
\begin{equation}
	\mathbf {y}^{\text {DD}} = \left(
	\left( \mathbf{W}_{\text{RF}} \mathbf{W}_{\text{BB}}\right)^{H}  \otimes 
	\left( {\mathbf {F}_{N} \otimes \mathbf {I}_{M} } \right) 
	\right) \mathbf {y}^{\text {TD}}.
	\label{equ23}
\end{equation}

As for both transmitter and receiver, we consider a uniform planar array (UPA) equipped with $N_y \times N_z$ elements along the $yz$-plane. The array response vector can be expressed as
\begin{align*}
	\mathbf {a}(\theta,\phi) &= \mathbf{a}_z(\phi) \otimes \mathbf{a}_y(\theta, \phi) 
	\\
	&= \frac {1}{\sqrt {N_{y}{N_{z}}}} [1, \ldots, \mathrm {e}^{j2\pi d(n_{y}\mathrm {sin}\theta \mathrm {sin}\phi \!+\!n_{z}\mathrm {cos}\phi)/\lambda }, \\
		& \ldots,\mathrm {e}^{j2\pi d((N_{y}-1)\mathrm {sin}\theta \mathrm {sin}\phi +\!(N_{z}-1)\mathrm {cos}\phi)/\lambda }]^{\mathrm {T}},\tag{23}
        \setcounter{equation}{23}
\end{align*}
where $\theta$ and $\phi$ refer to the azimuth and elevation angle, respectively.

Different from the array response matrix for communication, the sensing array response matrix can be defined as
\begin{equation}
	\mathbf {A}(\theta,\phi) = \mathbf {a}(\theta,\phi) \mathbf {a}^{T}(\theta,\phi),
\end{equation}
According to (\ref{equ10}), the single-input single-output time domain channel matrix can be given by
\begin{equation}
	\mathbf H^{\text {TD}}_{{\text {s}}} = \sum_{p=1}^P \alpha_p \mathbf{\Delta}_p \mathbf{G}_{l_p} \in \mathbb{C}^{MN \times MN},
\end{equation}
Thus the multiple-input multiple-output channel matrix $ \mathbf H^{\text {TD}}_{{\text {m}}}\in \mathbb{C}^{N_r MN \times N_t MN}$ is derived as
\begin{equation}
	\mathbf H^{\text {TD}}_{{\text {m}}} = \sqrt {{N_{t}}{N_{r}}} \sum_{p=1}^P \alpha_p \mathbf {A}(\theta_p,\phi_p) \otimes \mathbf{\Delta}_p \mathbf{G}_{l_p}.
	\label{equ27}
\end{equation}
The time domain receive vector in (\ref{equ23}) is given by
\begin{equation}
	\mathbf {y}^{\text {TD}} = \mathbf H^{\text {TD}}_{{\text {m}}} \mathbf {x}^{\text {TD}} + \mathbf {n}^{\text {TD}},
	\label{equ28}
\end{equation}
where $\mathbf {n}^{\text {TD}}$ is the zero-mean addictive noise vector. By plugging (\ref{equ27}), (\ref{equ28}) and (\ref{equ22}) into (\ref{equ23}), we have the equation at the top, as (\ref{equ29}).

For simplicity, we define $\mathbf {Y}^{\text {DD}} = \text{vec}^{-1} (\mathbf {y}^{\text {DD}})$. Based on the fundamental property of vectorization, (\ref{equ29}) can be rearranged as
\begin{align*}
	& \! \mathbf {Y}^{\text {DD}} \!=\! \sqrt {{N_{t}}{N_{r}}} \sum_{p=1}^P \alpha_p 
	\left( {\mathbf {F}_{N} \otimes \mathbf {I}_{M} }\right)  \mathbf{\Delta}_p \mathbf{G}_{l_p}
	\left( {\mathbf {F}_{N}^{H} \otimes \mathbf {I}_{M} }\right)
	\mathbf {X}^{\text {DD}}\\
	& \quad  \left( \mathbf{F}_{\text{RF}} \!\mathbf{F}_{\text{BB}} \right)^{T} \! \mathbf {A}^{T}\!(\theta_p,\phi_p)\! \left(\! \mathbf{W}_{\text{RF}}\! \mathbf{W}_{\text{BB}}\!\right)^{\ast} 
	\!+\! \mathbf{\tilde{N}}\! \left(\! \mathbf{W}_{\text{RF}} \mathbf{W}_{\text{BB}}\right)^{\ast}\!,\!
	\tag{29}
	\label{equ31}
\end{align*}
where $\mathbf{\tilde{N}}$ is the delay-Doppler domain noise matrix. We define $\mathbf Y = \left({\mathbf {F}_{N}^{H} \otimes \mathbf {I}_{M} }\right) \mathbf {Y}^{\text {DD}}$ and $\mathbf X = \left({\mathbf {F}_{N}^{H} \otimes \mathbf {I}_{M} }\right) \mathbf {X}^{\text {DD}}$, and thus (\ref{equ31}) is updated as
\begin{align*}
	\mathbf {Y} &= \sqrt {{N_{t}}{N_{r}}} \sum_{p=1}^P \alpha_p 
	\mathbf{\Delta}_p \mathbf{G}_{l_p} \mathbf {X}
	\left( \mathbf{F}_{\text{RF}} \mathbf{F}_{\text{BB}} \right)^{T} \mathbf {A}^{T}(\theta_p,\phi_p) \\
	& \left( \mathbf{W}_{\text{RF}} 
	\mathbf{W}_{\text{BB}}\right)^{\ast} 
	+ \mathbf{N} \left( \mathbf{W}_{\text{RF}} \mathbf{W}_{\text{BB}}\right)^{\ast},
	\tag{30}
	\label{equ32}
    \setcounter{equation}{30}
\end{align*}
where time domain noise matrix $\mathbf{N} \in \mathbb{C}^{N_t \times N_r}$ satisfies $\mathrm{vec}(\mathbf{N})\sim\mathcal{CN}(\mathbf{0},\sigma^2\mathbf{I})$. $\sigma^2$ refer to the noise variance.

\section{Multi-dimension Sensing Parameters Estimation Algorithm}\label{sec:sensing_algo}
In this section, we propose a multi-dimension estimation algorithm to estimate azimuth angle $\theta$, elevation angle $\phi$, delay $\tau$ and Doppler shift $\nu$. For simplicity, in the following section, we define $\mathbf{F} = \mathbf{F}_{\text{RF}} \mathbf{F}_{\text{BB}} \in \mathbb{C}^{N_t \times N_s}$ as the hybrid precoder matrix and $\mathbf{W} = \mathbf{W}_{\text{RF}} \mathbf{W}_{\text{BB}} \in \mathbb{C}^{N_r \times N_s}$ as the hybrid sensing combiner matrix. Once we have the hybrid beamforming matrix, the analog and digital beamformers can be calculated by well-studied algortithms, including alternating minimization algorithm~\cite{yu2016alternating}.
\subsection{Maximum Likelihood Estimator}
The derivation of maximum likelihood estimator (MLE) requires to minimize the log-likelihood function given by
\begin{equation} 
l\left ({\mathbf {Y}|\mathbf{X}, \theta,\phi,\tau,\nu}\right)= \left \lVert{ \mathbf {Y} - \alpha \mathbf{\Delta}(\nu) \mathbf{G}(\tau) \mathbf {X} \mathbf{F}^{T} \mathbf {A}^{T}(\theta,\phi) \mathbf{W}^{\ast}}
\right \rVert _{F}^{2},
\end{equation}
since the signal model can be represented as 
\begin{equation} 
\mathbf {Y} = \alpha \mathbf{\Delta}(\nu) \mathbf{G}(\tau) \mathbf {X} \mathbf{F}^{T} \mathbf {A}^{T}(\theta,\phi) \mathbf{W}^{\ast} +  \mathbf{N} \mathbf{W}^{\ast},
\end{equation}
Note that we focus on the single target case, since the estimator can be expanded to the multi-target case with another step to eliminate the interference from other targets. The estimator can be described as
\begin{align*}
	(\hat {\tau }, \hat {\nu }, \hat {\theta }, \hat {\phi },\hat {\alpha }) &= \arg \min _{(\tau, \nu,\theta, \phi,\alpha)} l\left ({\mathbf {Y}|\mathbf{X}, \theta,\phi,\tau,\nu}\right) \\
    &=\arg \min _{(\tau, \nu,\theta, \phi,\alpha)} \left \lVert{ \mathbf {Y} - \alpha \mathbf{B}(\tau, \nu,\theta, \phi)}
    \right \rVert _{F}^{2},
    \tag{33}
    \setcounter{equation}{33}
    \label{equ35}
\end{align*}
where $\mathbf{B}(\tau, \nu,\theta, \phi) = \mathbf{\Delta}(\nu) \mathbf{G}(\tau) \mathbf {X} \mathbf{F}^{T} \mathbf {A}^{T}(\theta,\phi) \mathbf{W}^{\ast}$ denotes an observation matrix corresponding to $(\tau, \nu,\theta, \phi)$ for the sake of simplicity.

Under the assumption that the variables $(\tau, \nu,\theta, \phi)$ are known as the condition for log-likelihood function, we can calculate the optimal channel coefficient $\alpha$ by letting the partial derivative on $\alpha$ as zero, we have 
\begin{equation}
    \hat{\alpha} = \frac{\text{tr} \left( \mathbf{Y}^{H} \mathbf{B}(\tau, \nu,\theta, \phi)\right)}{\left \lVert{  \mathbf{B}(\tau, \nu,\theta, \phi) }\right \rVert _{F}^{2}},
    \label{equ36}
\end{equation}
By substituting (\ref{equ36}) into (\ref{equ35}), the minimization problem boils down to
\begin{equation}
    (\hat {\tau }, \hat {\nu }, \hat {\theta }, \hat {\phi }) 
    = \arg \max _{(\tau, \nu,\theta, \phi)} 
    \frac{\left| \text{tr} \left(\mathbf{Y}^{H} \mathbf{B}(\tau, \nu,\theta, \phi)\right) \right|^{2}}{\left \lVert{  \mathbf{B}(\tau, \nu,\theta, \phi) }\right \rVert _{F}^{2}}.
    \label{equ37}
\end{equation}

\subsection{Estimation Algorithm Design}\label{estimation_algorithm}
To estimate unknown parameters of the target, i.e. $(\tau, \nu,\theta, \phi)$, we have to solve the maximization problem, given by (\ref{equ37}). However, optimization over four dimension variables $(\tau, \nu,\theta, \phi)$ suffers from explosive growth in sample complexity. To solve this, our algorithm narrows down the range of 4D search by introducing more steps of approximate estimation of low dimension. Specifically, we estimate azimuth angle $\theta$ and elevation angle $\phi$ in the 2D spatial domain, and then estimate the delay $\tau$ and Doppler shift $\nu$ in the delay-Doppler domain. Both results from 2D spatial domain and delay-Doppler domain are approximate results, which will be further refined with (\ref{equ37}).

\textit{1) Approximate azimuth and elevation angle estimation with MUSIC algorithm.}

To reduce the complexity of searching the entire 4D area, we consider to first obtain an approximate estimation of azimuth and elevation angles with multiple signal classification (MUSIC) algorithm~\cite{hayes1996statistical}. The covariance matrix is given by
\begin{equation}
    \mathbf{R}_{y} = \frac{1}{M N} \mathbf{Y}^{T} \left(\mathbf{Y}^{T}\right)^{H}.
\end{equation}
Then we conduct eigenvalue decomposition (EVD) as
\begin{equation}
	\mathbf{R}_{y} = \mathbf{U}_s \mathbf{\Sigma}_s \mathbf{U}_s^H + \mathbf{U}_n \mathbf{\Sigma}_n \mathbf{U}_n^H,
\end{equation}
where $\mathbf{\Sigma}_s \in \mathbb{C}^{L \times L}$ consists of $L$ leading eigenvalues, $\mathbf{\Sigma}_n \in \mathbb{C}^{(N_\text{RF}^r - L) \times (N_\text{RF}^r - L)}$ contains the remaining eigenvalues. Then we can estimate azimuth and elevation angles by
\begin{equation}
   (\hat {\theta }, \hat {\phi }) = \arg \max_{(\theta, \phi)}  \frac{\mathbf{a}^H(\theta, \phi) \mathbf{W} \mathbf{W}^H \mathbf{a}(\theta, \phi)}{\mathbf{a}^H(\theta, \phi) \mathbf{W} \mathbf{U}_n \mathbf{U}_n^H \mathbf{W}^H \mathbf{a}(\theta, \phi)}.
\end{equation}

\textit{2) Approximate delay and Doppler shift estimation with ML algorithm.}

Given the estimated angle information, we can estimate delay and Doppler shift with
\begin{equation}
    (\hat {\tau }, \hat {\nu }) = \arg \max_{(\tau, \nu)} \frac{\left| \text{tr} \left(\mathbf{Y}^{H} \mathbf{\Delta}(\nu) \mathbf{G}(\tau) \mathbf {\hat X}\right) \right|^{2}}{\left \lVert{  \mathbf{\Delta}(\nu) \mathbf{G}(\tau) \mathbf {\hat X} }\right \rVert _{F}^{2}},
\end{equation}
where 
\begin{equation}
    \mathbf {\hat X} = \mathbf {X} \mathbf{F}^{T} \mathbf {A}^{T}(\hat {\theta }, \hat {\phi }) \mathbf{W}^{\ast}.
\end{equation}

In this estimation procedure, we reduce the sample complexity by first performing on-grid search followed by off-grid search to narrow the region of uncertainty that contains the true values of $\tau$ and $\nu$.

\textit{3) Joint estimation for azimuth angle, elevation angle, path delay and path Doppler shift.}

The first two steps of approximate estimation provide relatively accurate estimation results of $(\tau, \nu,\theta, \phi)$. Thus, the last step of our algorithm is to further improve the estimation accuracy by the joint estimation of four variables within a much smaller search space, given by
\begin{equation}
    (\hat {\tau }, \hat {\nu }, \hat {\theta }, \hat {\phi }) 
    = \arg \max _{(\tau, \nu,\theta, \phi)} 
    \frac{\left| \text{tr} \left(\mathbf{Y}^{H} \mathbf{B}(\tau, \nu,\theta, \phi)\right) \right|^{2}}{\left \lVert{  \mathbf{B}(\tau, \nu,\theta, \phi) }\right \rVert _{F}^{2}}.
\end{equation}

Our estimation algorithm not only reduces its sample complexity by designing efficient search strategy over four dimension variables, but also reduces the computational complexity by exploiting the property of circular matrix. Specifically, the multiplication of a circular matrix $\mathbf{G}(\tau)$ and a $MN$-length vector $\mathbf{x}$ is equivalent to shifted circular convolution of $\mathbf {x}$ and $(2Q+1)$-length raised cosine filter $\mathbf {g}$, where $\mathbf {g} = \left\{g(q T_s)\right\}_{-Q \leq q \leq Q}$, given by
\begin{equation}
	\mathbf{G}(\tau) \mathbf {x} = \boldsymbol{\Pi}_{MN}^{\lfloor \frac{\tau}{T_s} \rfloor}(\mathbf{g} \circledast \mathbf{x}),
    \label{equ44}
\end{equation}
where $\circledast$ denotes the circular convolution operation. Since circular convolution in time domain is equivalent to the multiplication of the Fourier coefficients in the frequency domain, the complexity of (\ref{equ44}) is reduced to $\mathcal{O}\left(MN\operatorname{log}(MN)\right)$.
\subsection{Cram\'er Rao Lower Bound}\label{sensing_CRLB}
Cram\'er Rao lower bounds provide a theoretical lower bound on the variance of any unbiased estimator for a parameter. The unknown parameters are azimuth angle $\theta$, elevation angle $\phi$, delay $\tau$ and Doppler shift $\nu$, that is $\boldsymbol{\xi} = [\theta,\phi,\tau,\nu]$. The main diagonal elements of the inverse of the Fisher information matrix (FIM). Each FIM element can be calculated by \cite{song_intelligent_2023}
\begin{equation} 
	\begin{split} 
		 \! {\mathbf {J}}_{\!\xi _{i} \xi _{j}\!}  \! = 
		2\text{Re}\!\left\lbrace \! \frac{\partial {\mathbf {y}}^{H}}{\partial \bm \xi _{i}}\mathbf {R}_{n}^{-1}\frac{\partial {\mathbf {y}}}{\partial \bm \xi _{j}}\!\right\rbrace
		\!+\!
		\text{tr}\!\left(\!\mathbf {R}_{n}^{-1}\!\frac{\partial \mathbf {R}_{n}}{\partial \bm \xi _{i}}\mathbf {R}_{n}^{-1}\!\frac{\partial \mathbf {R}_{n}}{\partial \bm \xi _{j}}\!\right)\!,
	\end{split}
    \label{equ45}
\end{equation}
where $\mathbf{y} = \text{vec}(\mathbf{Y})$. The noise covariance matrix $\mathbf {R}_{n}$ can be calculated as
\begin{align*}
    \mathbf {R}_{n} &= \frac{1}{M N N_{s}} \text{vec}(\mathbf{N}  \mathbf{W}^{\ast}) \text{vec}(\mathbf{N} \mathbf{W}^{\ast})^{H} \\
    &= \frac{1}{M N N_{s}} \left( \mathbf{W}^{H} \otimes \mathbf{I}_{MN}\right) \text{vec}(\mathbf{N})  \text{vec}(\mathbf{N})^{H} \left( \mathbf{W} \otimes \mathbf{I}_{MN}\right) \\
    &= \sigma^2 \left( \mathbf{W}^{H} \mathbf{W} \otimes \mathbf{I}_{MN}\right),
    \tag{44}
	\setcounter{equation}{44}    
	\label{equ46}
\end{align*}
Thus, the FIM element can be updated as
\begin{equation} 
		\!{\mathbf {J}}_{\xi _{i} \xi _{j}} \!= 
		\!\frac{2}{\sigma^2}\text{Re}\left\lbrace \!
        \left(\! \frac{\partial}{\partial \bm \xi _{i}}\mathbf {y}\right)^{H}\!
        \left( \mathbf{W}^{H}\! \mathbf{W} \!\otimes\! \mathbf{I}_{MN}\right)^{-1}
        \frac{\partial}{\partial \bm \xi _{j}} \mathbf {y}\right\rbrace.
    \label{equ47}
\end{equation}

Due to space limitations, the detailed derivation of FIM elements is omitted. We observe that, sensing accuracy is theoretically dependent on the noise covariance. It implies that sensing pays more attention on the noise after the procedure of combining while communication often ignores its influence and regards the noise covariance as identity matrix. Thus, combiner design for sensing is much more important than that for communication. 

\section{Joint hybrid beamforming design for sensing and communication}\label{sec:beamforming_design}
We consider a transmission framework as shown in Fig. \ref{fig: concept diagram}, there are three hybrid beamformers of interests, including, $(i)$ the precoder at the base station whose signal is processed for both sensing and communication, $(ii)$ the sensing combiner co-located at the base station to process the echo signal reflected by the target, and $(iii)$ the communication combiner at the user equipment to demodulate the signal. 

Sec. \ref{sensing_CRLB} demonstrates that the sensing combiner can affect the sensing accuracy by influencing the noise covariance matrix. There are strict restrictions on the design of sensing combiner. A basic restriction is that $\mathbf{W}^{H} \mathbf{W}$ should be of full rank, according to (\ref{equ47}). In light of this, we endeavor to optimize the design of sensing combiner. After designing the sensing combiner by optimizing sensing ability, then we consider to design the precoder to achieve both the functionalities of sensing and communication.

\subsection{Problem Formulation}
There are two problems considering two distinct metrics, for sensing and communication, respectively. Firstly, since the beamforming design only functions in the spatial domain, we consider the CRLB for azimuth angle and elevation angle as the sensing metric. The CRLB minimization problem to optimize the sensing performance is formulated as
\begin{equation} 
	{\begin{array}{c}\mathop {\min }\limits _{\mathbf{W}, \mathbf{F}} \;\operatorname{CRLB} \left(\theta \right) + \operatorname{CRLB} \left(\phi \right)
	\\ \;\; \text{s.t.}\;\; \|\mathbf{W}\|_F^2 = N_s, \|\mathbf{F}\|_F^2 = N_s.\\ \end{array}} 
    \label{equ48}
\end{equation}

On the other hand, spectral efficiency is a common metric to optimize the communication performance, given by 
\begin{equation}
     {\begin{array}{c}\mathop {\max }\limits _{\mathbf{F},\mathbf{C}} \; 
     \log \det\bigg(\mathbf{I}_{N_s} + \frac{\rho}{N_s} \mathbf{R}_n^{-1} \mathbf{C}^H \mathbf{H}_c \mathbf{F} \mathbf{F}^H   \mathbf{H}_c^H \mathbf{C}\bigg)
	\\ \;\;\;\;\; \text{s.t.}\;\; \|\mathbf{F}\|_F^2 = N_s, \|\mathbf{C}\|_F^2 = N_s.\\
    \end{array}}    
    \label{equ49}
\end{equation}
where $\mathbf{C}$ is the communication hybrid combiner. $\rho$ denotes the transmit power and $\mathbf{R}_n = \sigma_n^2 \mathbf{C}^H \mathbf{C}$ is a noise covariance matrix of the communication channel. Generally, the optimal solution to (\ref{equ49}) is the singular value decomposition (SVD) of the channel.

We can observe that both two problems are related to the design of precoder, which is the reason why most existing studies focus on the transmit design to achieve both functionalities of sensing and communication. Instead, as we illustrate the importance of sensing combiner before, we first optimize the sensing combiner with prerequisite precoder to minimize the CRLB of angles. Afterwards, based on optimized sensing combiner and its corresponding sensing performance, we further involve the optimization of communication performance. Specifically, we design the precoder, to achieve higher spectral efficiency while, to some extent, maintaining its sensing accuracy.

\subsection{Cram\'er-Rao Lower Bound Derivation}\label{CRLBderive}
In this subsection, we derive the CRLBs of target angles to further illustrate how precoder and combiner affect sensing and as a metric for beamforming design. (\ref{equ32}) can be simplified as 
\begin{equation}
	\mathbf{R} = \alpha \mathbf{W}^{H} \mathbf{A}(\theta,\phi) \mathbf{F} \mathbf{S} + \mathbf{W}^{H} \mathbf{Z},
\end{equation}
where $\mathbf{R} =  \left[ \mathbf {Y} \right]^{T} $ denotes the receive matrix, and $\mathbf{S} = \left[ \mathbf {X} \right]^{T} [\mathbf{G}_{l}]^{T} \mathbf{\Delta} $ denotes the equivalent transmit matrix. Noise matrix $\mathbf{Z} =  \left[ \mathbf {N} \right]^{T} $ satisfies the condition that $\mathrm{vec}(\mathbf{N}^{T})\sim\mathcal{CN}(\mathbf{0},\sigma^2\mathbf{I})$. We assume that the sample covariance matrix is an identity matrix, given by
\begin{equation}
	\mathbf{R}_{x} = \frac{1}{M N N_s}\text {vec}(\mathbf{X}) \text {vec}(\mathbf{X})^{H} = \mathbf{I}_{M N N_s}.
\end{equation}
Then the covariance matrix of $\mathbf{S}$ can be calculated as
\begin{align*}
	\mathbf{R}_{s} &= \frac{1}{M N N_s}\text {vec}(\mathbf{S}) \text {vec}(\mathbf{S})^{H} 
	\\ &= \frac{1}{M N N_s} \left( \mathbf{\Delta} \mathbf{G}_{l} \otimes \mathbf{I} \right) \text {vec}(\mathbf{X}^{T}) \text {vec}(\mathbf{X}^{T})^{H} \left( \mathbf{G}_{l}^{H} \mathbf{\Delta}^{*}  \otimes \mathbf{I} \right)
	\\ & = \left( \mathbf{\Delta} \mathbf{G}_{l} \otimes \mathbf{I} \right) \mathbf{P} \mathbf{R}_{x} \mathbf{P}^{H}
	\left( \mathbf{G}_{l}^{H} \mathbf{\Delta}^{*}  \otimes \mathbf{I} \right)
	\\ & \approx \left( \mathbf{\Delta} \mathbf{G}_{l} \mathbf{G}_{l}^{H} \mathbf{\Delta}^{*}\right) \otimes \mathbf{I}_{N_s},
    \tag{50}
\end{align*}
where $\mathbf{P}$ is the permutation matrix which satisfies $\mathbf{P}\mathbf{P}^{H}= \mathbf{I}$. Besides, $\mathbf{G}_{l}$ is a circular matrix which satisfies $\mathbf{G}_{l} \mathbf{G}_{l}^{H} \approx \mathbf{I}$.

The noise covariance matrix, denoted as $\mathbf{R}_{n}$, is given by
\begin{align*}
    \mathbf{R}_{n} &= \frac{1}{M N N_s}\text {vec}({\mathbf{W}^{H} \mathbf{Z}}) \text {vec}({\mathbf{W}^{H} \mathbf{Z}})^{H}  \\
    &= \frac{1}{M N N_s} \left(\mathbf{I} \otimes \mathbf{W}^{H}\right) \text {vec}(\mathbf{Z}) \text {vec}(\mathbf{Z})^{H} \left(\mathbf{I} \otimes \mathbf{W}\right) \\
    &= \sigma^2 \left[ \mathbf{I}_{MN} \otimes \left( \mathbf{W}^{H} \mathbf{W}\right) \right] 
    ,\tag{51}
    \label{equ53}
    \setcounter{equation}{51}
\end{align*}

Since we only consider the CRLB of angles, the FIM can be expressed as follows for simplicity.
\begin{equation} 
	{\mathbf J}  = 
	\left[ 
	{{\begin{array}{ccc} 
				{{{\mathbf J}}_{{{ {\theta {\theta}} } }}} &{{{{\mathbf J}}_{{{\theta \phi }}}}} & {{{{\mathbf J}}_{{\theta {\boldsymbol \alpha} }}}}\\ 
				{{{{\mathbf J}}_{{{\phi \theta }}}}}&{{{{\mathbf J}}_{{{\phi \phi }}}}}&{{{{\mathbf J}}_{{\phi \boldsymbol{\alpha }}}}}\\
				{{{{\mathbf J}}_{{\boldsymbol{\alpha} \theta }}}}&{{{{\mathbf J}}_{\boldsymbol{\alpha} \phi}}}&{{{{\mathbf J}}_{{\boldsymbol{\alpha \alpha }}}}} 
	\end{array}}} 
	\right],
\end{equation}
where each FIM element can be calculated by (\ref{equ45}).
\begin{figure*}[!t] 
	\normalsize
	\begin{align*}
	f_{\bm \xi _{i} \bm \xi _{j}}=\left\{
\begin{array}{rcl}
\text{Re}\left\lbrace \text{tr} \left( \mathbf{F}^{H} \dot{\mathbf{A}}^{H}_{\bm \xi _{i}} \mathbf{W} \left( \! \mathbf{W}^{H} \mathbf{W} \right)\!^{-1} \mathbf{W}^{H} \dot{\mathbf{A}}_{\bm \xi _{j}} \mathbf{F} \right) \right\rbrace
    &      & {\bm \xi _{i},\bm \xi _{j} \in \lbrace \theta,\phi\rbrace}
\\
\text{Re} \left\lbrace \text{tr} \left(  \mathbf{F}^{H} \mathbf{A}^{H} \mathbf{W} \left( \mathbf{W}^{H} \mathbf{W}\right)^{-1} \mathbf{W}^{H} \mathbf{A} \mathbf{F} \right) \right\rbrace     &      & {\bm \xi _{i}=\bm \xi _{j} = {\bm \alpha}}\\
\text{Re} \! \left\lbrace 
	\text{tr} \! \left(  \mathbf{F}^{H} \! \dot{\mathbf{A}}^{H}_{\bm \xi _{i}} \mathbf{W} \left( \mathbf{W}^{H} \mathbf{W}\right)\!^{-1} \mathbf{W}^{H} \! \mathbf{A} \mathbf{F} \right) \right\rbrace   
 &      & {\bm \xi _{i} \in \lbrace \theta,\phi\rbrace , \bm \xi _{j} = {\bm \alpha}}
\end{array} \right.
	\tag{75}
    \label{equ79}
	\end{align*}
    \hrulefill
\end{figure*}

The noise covariance matrix $\mathbf{R}_{n}$ is independent of $\bm \xi$. Thus, we have $\frac{\partial \mathbf {R}_{n}}{\partial \bm \xi _{i}} = 0$. In (\ref{equ45}), $\mathbf {h}$ can be written as
\begin{equation}
	\mathbf {h} = \alpha \text{vec}\left( \mathbf{W}^{H} \mathbf{A}(\theta,\phi) \mathbf{F} \mathbf{S}\right),
\end{equation}
The partial derivative of $\mathbf {h}$ can be expressed as 
\begin{equation}
	\frac{\partial {\mathbf {h}}}{\partial \theta} = \alpha \text{vec}\left( \mathbf{W}^{H} \dot{\mathbf{A}}_{\theta} \mathbf{F} \mathbf{S}\right),
	\label{equ56}
\end{equation}
\begin{equation}
	\frac{\partial {\mathbf {h}}}{\partial \phi} = \alpha \text{vec}\left( \mathbf{W}^{H} \dot{\mathbf{A}}_{\phi} \mathbf{F} \mathbf{S}\right),
\end{equation}
\begin{equation}
	\frac{\partial {\mathbf {h}}}{\partial {\boldsymbol {\alpha}}} =  [1,j] \otimes \text{vec}\left( \mathbf{W}^{H} \mathbf{A} \mathbf{F} \mathbf{S}\right),
	\label{equ58}
\end{equation}	
where $\dot{\mathbf{A}}_{\theta}$ and $\dot{\mathbf{A}}_{\phi}$ denote the partial derivative of $\mathbf{A}$ with respect to $\theta$ and $\phi$ respectively.  The FIM elements are listed as follows.
\begin{equation}
	\!{\mathbf J}_{{ {\theta {\theta}}}} \!= \! \frac{2 \beta |\alpha |^{2}}{\sigma^2}\!   \text{Re}\!\left\lbrace
	\text{tr}\!\left( \mathbf{F}^{H} \!\dot{\mathbf{A}}^{H}_{\theta} \mathbf{W}\! \left( \mathbf{W}^{H} \!\mathbf{W} \right)^{-1} \mathbf{W}^{H} \!\dot{\mathbf{A}}_{\theta} \mathbf{F} \!
	\right) \!
	\right\rbrace \!,\!
	\label{equ59}
\end{equation}
\begin{equation}
	\!{\mathbf J}_{{ {\phi {\phi}}}} \!= \!  \frac{2 \beta |\alpha |^{2}}{\sigma^2}\!  \text{Re}\!\left\lbrace
	\text{tr}\!\left( \mathbf{F}^{H} \!\dot{\mathbf{A}}^{H}_{\phi} \mathbf{W}\!\left( \mathbf{W}^{H} \!\mathbf{W} \right)^{-1} \mathbf{W}^{H} \!\dot{\mathbf{A}}_{\phi} \mathbf{F} \!
	\right) \!
	\right\rbrace \!,\!
\end{equation}
\begin{equation} 
	\!{\mathbf J}_{{ {\theta {\phi}}}} \!= \!  \frac{2 \beta |\alpha |^{2}}{\sigma^2}\! \text{Re}\!\left\lbrace
	\text{tr}\!\left(  \mathbf{F}^{H} \!\dot{\mathbf{A}}^{H}_{\theta} \mathbf{W} \!\left( \mathbf{W}^{H} \!\mathbf{W}\right)^{-1} \mathbf{W}^{H} \!\dot{\mathbf{A}}_{\phi} \mathbf{F} \!
	\right) \!
	\right\rbrace \!,\!
\end{equation}
\begin{equation}
	\!{\mathbf J}_{{\boldsymbol {\alpha \alpha}}}  \!= \! \frac{2 \beta}{\sigma^2} \!
    \text{Re}\!\left\lbrace 
    \text{tr}\!\left(  \mathbf{F}^{H} \!\mathbf{A}^{H} \mathbf{W} \!\left( \mathbf{W}^{H} \!\mathbf{W}\right)^{-1} \mathbf{W}^{H} \!\mathbf{A} \mathbf{F} \!
    \right) \!
    \right\rbrace \mathbf{I}_{2},\!
\end{equation}
\begin{equation}
	\!{\mathbf J}_{{\theta \boldsymbol {\alpha}}}  \!= \! \frac{2 \beta}{\sigma^2} \!
    \text{Re} \! \left\lbrace 
	\text{tr} \! \left(  \mathbf{F}^{H} \! \dot{\mathbf{A}}^{H}_{\theta} \!\mathbf{W} \!\left( \mathbf{W}^{H} \!\mathbf{W}\right)^{-1} \!\mathbf{W}^{H} \! \mathbf{A} \!\mathbf{F} \!\right) \! \alpha^{*} \! [1,j]\!
	\right\rbrace \!,\!
\end{equation}
\begin{equation}
	\!{\mathbf J}_{{\phi \boldsymbol {\alpha}}} \!= \! \frac{2 \beta}{\sigma^2} \!
    \text{Re} \! \left\lbrace 
	\text{tr} \! \left(  \mathbf{F}^{H} \! \dot{\mathbf{A}}^{H}_{\phi} \!\mathbf{W} \!\left( \mathbf{W}^{H} \!\mathbf{W}\right)^{-1} \!\mathbf{W}^{H} \! \mathbf{A} \!\mathbf{F} \right) \! \alpha^{*} \! [1,j] \!
	\right\rbrace\!,\!
\end{equation}
where $\beta = M N N_s \|\mathbf{\Delta} \mathbf{G}_{l}\|_F^2$. The detailed derivation of FIM element is in Appendix A. 

Based on the theoretical results, we can analyze how precoder and sensing combiner influence the sensing accuracy. According to (\ref{equ59}), the precoder $\mathbf{F}$ governs the transmit beam pattern, thereby affecting the sensing accuracy by determining the exact signal power directed toward the target. By comparison, the sensing combiner $\mathbf{W}$ determines the projection matrix $\mathbf{W} \left( \mathbf{W}^{H} \mathbf{W}\right)^{-1} \mathbf{W}^{H}$ and modifies sensing accuracy by determining the projected received power in the subspace spanned by $\mathbf{W}$.

Since the steering vector $\mathbf {a}(\theta,\phi) = \mathbf{a}_z(\phi) \otimes \mathbf{a}_y(\theta, \phi)$ can be decomposed by the kronecker product of two vectors, given by
\begin{equation}
	\mathbf{a}_z(\phi) = \frac {1}{\sqrt {{N_{z}}}} \left[ 1, \ldots, 
	\mathrm {e}^{j \frac{2\pi d}{\lambda} n_{z}\mathrm {cos}\phi}, \ldots
	\right]^{T},
\end{equation}
\begin{equation}
	\mathbf{a}_y(\theta, \phi) = \frac {1}{\sqrt {{N_{y}}}} \left[ 1, \ldots, 
	\mathrm {e}^{j \frac{2\pi d}{\lambda} n_{y}\mathrm {sin}\theta \mathrm {sin}\phi}, \ldots
	\right]^{T}.
\end{equation}
In this case, the partial derivative of $\mathbf{A}(\theta,\phi)= \mathbf{a}(\theta,\phi) \mathbf{a}^{T}(\theta,\phi)$ can be calculated as
\begin{equation}
	\dot{\mathbf{A}}_{\theta} = \frac{\partial {\mathbf {a}}}{\partial \theta} \mathbf{a}^{T} + \mathbf{a} \left( \frac{\partial {\mathbf {a}}}{\partial \theta}\right) ^{T},
\end{equation}
\begin{equation}
	\dot{\mathbf{A}}_{\phi} = \frac{\partial {\mathbf {a}}}{\partial \phi} \mathbf{a}^{T} + \mathbf{a} \left( \frac{\partial {\mathbf {a}}}{\partial \phi}\right) ^{T},
\end{equation}
where the partial derivative of $\mathbf{a}$ are given by
\begin{equation}
	\frac{\partial {\mathbf {a}}}{\partial \theta} = \left[ 
	j \frac{2\pi d}{\lambda} \mathrm {cos}\theta \mathrm {sin}\phi
	\right] \mathbf{a}_z(\phi) \otimes  \mathbf{D}_{y} \mathbf{a}_y(\theta, \phi),
\end{equation}
\begin{align*}
	\frac{\partial {\mathbf {a}}}{\partial \phi} = & \left[ 
	j \frac{2\pi d}{\lambda} (-\mathrm {sin}\phi) \right] 
	\mathbf{D}_{z} \mathbf{a}_z(\phi) \otimes   \mathbf{a}_y(\theta, \phi) 
    + \\
    & \left[j \frac{2\pi d}{\lambda} \mathrm {sin}\theta \mathrm {cos}\phi \right]
    \mathbf{a}_z(\phi) \otimes \mathbf{D}_{y} \mathbf{a}_y(\theta, \phi),
    \tag{68}
\end{align*}
where $\mathbf{D}_{z} = \text{diag}([0,1,\ldots,N_z-1])$ and $\mathbf{D}_{y} = \text{diag}([0,1,\ldots,N_y-1])$.

Finally, the CRLB for $\theta$ and $\phi$ can be written as
\begin{align*} 
	&(\operatorname{CRLB}(\theta))^{ - 1} 
	= {{{\mathbf J}}_{{ {\theta {\theta}}}}} 
	- {{{\mathbf J}}_{{{\theta {\boldsymbol \alpha} }}}}{{\mathbf J}}_{{\boldsymbol{\alpha \alpha }}}^{ - 1}{{\mathbf J}}_{{{\theta {\boldsymbol \alpha} }}}^T - 
	\left({{{{\mathbf J}}_{{{\theta \phi }}}} - {{{\mathbf J}}_{{{\theta {\boldsymbol \alpha} }}}}{{\mathbf J}}_{{\boldsymbol{\alpha \alpha }}}^{ - 1}{{\mathbf J}}_{{{\phi {\boldsymbol \alpha} }}}^T} \right)
	\nonumber
	\\
	&  \times
	{\left({{{{\mathbf J}}_{{{\phi \phi }}}} - {{{\mathbf J}}_{{{\phi {\boldsymbol \alpha} }}}}{{\mathbf J}}_{{\boldsymbol{\alpha \alpha }}}^{ - 1}{{\mathbf J}}_{{{\phi {\boldsymbol \alpha} }}}^T} \right)^{ - 1}}
	\left({{{{\mathbf J}}_{{{\phi \theta }}}} - {{{\mathbf J}}_{{{\phi {\boldsymbol \alpha} }}}}{{\mathbf J}}_{{\boldsymbol{\alpha \alpha }}}^{ - 1}{{\mathbf J}}_{{{\theta {\boldsymbol \alpha} }}}^T} \right).
    \tag{69}
\end{align*}
\begin{align*} 
	&(\operatorname{CRLB}(\phi))^{ - 1} 
	= {{{{\mathbf J}}_{{{\phi \phi }}}} - {{{\mathbf J}}_{{{\phi {\boldsymbol \alpha} }}}}{{\mathbf J}}_{{\boldsymbol{\alpha \alpha }}}^{ - 1}{{\mathbf J}}_{{{\phi {\boldsymbol \alpha} }}}^T}
	- \left({{{{\mathbf J}}_{{{\phi \theta }}}} - {{{\mathbf J}}_{{{\phi {\boldsymbol \alpha} }}}}{{\mathbf J}}_{{\boldsymbol{\alpha \alpha }}}^{ - 1}{{\mathbf J}}_{{{\theta {\boldsymbol \alpha} }}}^T}
	\right)
	\nonumber
	\\
	&  \times
	{\left({{{\mathbf J}}_{{ {\theta {\theta}}}}} 
		- {{{\mathbf J}}_{{{\theta {\boldsymbol \alpha} }}}}{{\mathbf J}}_{{\boldsymbol{\alpha \alpha }}}^{ - 1}{{\mathbf J}}_{{{\theta {\boldsymbol \alpha} }}}^T  \right)^{ - 1}}
	\left(
	{{{{\mathbf J}}_{{{\theta \phi }}}} - {{{\mathbf J}}_{{{\theta {\boldsymbol \alpha} }}}}{{\mathbf J}}_{{\boldsymbol{\alpha \alpha }}}^{ - 1}{{\mathbf J}}_{{{\phi {\boldsymbol \alpha} }}}^T} 
	\right). 
    \tag{70}
    \setcounter{equation}{70}
\end{align*}
Thereinto, some components can be calculated as
\begin{equation}
    {{{\mathbf J}}_{{ {\theta {\theta}}}}} 
	- {{{\mathbf J}}_{{{\theta {\boldsymbol \alpha} }}}}{{\mathbf J}}_{{\boldsymbol{\alpha \alpha }}}^{ - 1}{{\mathbf J}}_{{{\theta {\boldsymbol \alpha} }}}^T = \frac{2 \beta |\alpha |^{2}}{\sigma^2} 
    ({ f_{\theta \theta}}- \frac{f_{\theta {\boldsymbol \alpha}} f_{\theta {\boldsymbol \alpha}}}{f_{{\boldsymbol \alpha} {\boldsymbol \alpha}}}) = \frac{2 \beta |\alpha |^{2}}{\sigma^2} j_\theta
\end{equation}
\begin{equation}
    {{{\mathbf J}}_{{ {\theta {\phi}}}}} 
	- {{{\mathbf J}}_{{{\theta {\boldsymbol \alpha} }}}}{{\mathbf J}}_{{\boldsymbol{\alpha \alpha }}}^{ - 1}{{\mathbf J}}_{{{\phi {\boldsymbol \alpha} }}}^T 
    = \frac{2 \beta |\alpha |^{2}}{\sigma^2} 
    ({ f_{\theta \phi}}- \frac{f_{\theta {\boldsymbol \alpha}} f_{\phi {\boldsymbol \alpha}}}{f_{{\boldsymbol \alpha} {\boldsymbol \alpha}}})= \frac{2 \beta |\alpha |^{2}}{\sigma^2} j_{\theta\phi}
\end{equation}
\begin{equation}
    {{{\mathbf J}}_{{ {\phi {\phi}}}}} 
	- {{{\mathbf J}}_{{{\phi {\boldsymbol \alpha} }}}}{{\mathbf J}}_{{\boldsymbol{\alpha \alpha }}}^{ - 1}{{\mathbf J}}_{{{\phi {\boldsymbol \alpha} }}}^T 
    = \frac{2 \beta |\alpha |^{2}}{\sigma^2} 
    ({ f_{\phi \phi}}-
    \frac{f_{\phi {\boldsymbol \alpha}} f_{\phi {\boldsymbol \alpha}}}{f_{{\boldsymbol \alpha} {\boldsymbol \alpha}}}
    )= \frac{2 \beta |\alpha |^{2}}{\sigma^2} j_{\phi}
\end{equation}
\begin{equation}
    {{{\mathbf J}}_{{\phi}\theta }} 
	- {{{\mathbf J}}_{{{\phi {\boldsymbol \alpha} }}}}{{\mathbf J}}_{{\boldsymbol{\alpha \alpha }}}^{ - 1}{{\mathbf J}}_{{{\theta {\boldsymbol \alpha} }}}^T 
    = \frac{2 \beta |\alpha |^{2}}{\sigma^2} 
    ({ f_{\phi \theta}}-
    \frac{f_{\phi {\boldsymbol \alpha}} f_{\theta {\boldsymbol \alpha}}}{f_{{\boldsymbol \alpha} {\boldsymbol \alpha}}}
    )= \frac{2 \beta |\alpha |^{2}}{\sigma^2} j_{\phi\theta}
\end{equation}
where $j_{\bm \xi _{i}\bm \xi _{j}} = f_{\bm\xi_{i}\bm\xi_{j}} - f_{\bm \xi _{i}\boldsymbol \alpha}f_{\bm \xi _{j}\boldsymbol \alpha} / f_{\boldsymbol \alpha \boldsymbol \alpha}$ and $f_{\bm \xi _{i},\bm \xi _{j}}$ is represented as (\ref{equ79}). In this case, the CRLB value can be updated as 
\begin{equation}
\setcounter{equation}{76}
(\operatorname{CRLB}(\theta))^{ - 1} 
= \frac{2 \beta |\alpha |^{2}}{\sigma^2} 
    \left[ j_\theta - \frac{j_{\phi\theta}j_{\theta\phi}}{j_\phi}\right],
\end{equation}
\begin{equation}
(\operatorname{CRLB}(\phi))^{ - 1} 
= \frac{2 \beta |\alpha |^{2}}{\sigma^2} 
    \left[ j_\phi - \frac{j_{\theta\phi}j_{\phi\theta}}{j_\theta}\right].
\end{equation}
The problem in (\ref{equ48}) boils down to
\begin{equation} 
	{\begin{array}{c}\mathop {\min }\limits _{\mathbf{W}} \;  \frac{1}{j_\theta - \frac{j_{\phi\theta}j_{\theta\phi}}{j_\phi}} + \frac{1}{j_\phi - \frac{j_{\theta\phi}j_{\phi\theta}}{j_\theta}}
	\\ \;\;\;\;\; \text{s.t.}\;\; \|\mathbf{W}\|_F^2 = N_s, \\ \end{array}} 
    \label{equ41}
\end{equation}

\subsection{Sensing Combiner Design: CRLB Minimization}\label{CRLB_opt_algo}
We can observe that in (\ref{equ59}), calculating CRLB requires the information of target azimuth and elevation angle. However, these information are usually unknown, especially when the task is to estimate parameters of the target. To solve this problem, we can assume arbitrary azimuth and elevation angle to optimize an initial sensing combiner matrix, which will be regenerated to sweep the area according to beam scanning angle. Let $\bar \theta$ and $\bar \phi$ denote the assumed azimuth and elevation angle of the target. In this case, the optimal precoder for sensing can be set as \cite{elbir_terahertz-band_2021,wu_time-frequency-space_2024}
\begin{equation}
	\mathbf{F}_{s} = \frac{1}{\sqrt{N_t}} \mathbf {a}({\bar \theta},{\bar \phi}) \mathbf{1}_{N_s}^T,
	\label{equ83}
\end{equation}
which steers the beam of all RF chains towards the target. With preset precoder, we can optimize sensing combiner. Inspired by the fact that the optimal beamforming vector is the steering vector corresponding to the direction of interest, we develop a genetic-based optimization algorithm to solve (\ref{equ41}). 
\begin{itemize}
    \item Generate a random initial population. The algorithm starts by generating a random initial population of size $P_s$, each of which is a ${N_r \times N_\text{RF}}$ combining matrix, which is regarded as the parent generation.
    \item Survival of the fittest. Each member of the population is evaluated based on its fitness, given by $\left[ \operatorname{CRLB} \left(\theta \right) + \operatorname{CRLB} \left(\phi \right)\right]$. Individuals with lower fitness are selected as elites, with only these elite individuals surviving to contribute to the next generation. The ratio of elite individuals to the whole population is termed as elite rate.
    \item Create new generation. A new population is generated based on the elites from the previous generation. Each individual in the population consists of $N_\text{RF}$ columns, which are treated as $N_\text{RF}$ genes. New individuals are produced by performing crossover operations on the genes of selected parent individuals.
    \item Iterate the above procedure until convergence. The individual with the lowest fitness is the optimized solution.
\end{itemize}
\begin{table}[t]
    \centering
    \caption{Simulation Parameters}
    \label{tab:parameter}
    \begin{tabular}{ccc}
    \toprule
    \textbf{Notations} & \textbf{Definition} & \textbf{Value} \\
    \midrule
    $f_c$ & Carrier frequency & 0.3 THz~\cite{wu_time-frequency-space_2024} \\
    $\Delta f$ & Subcarrier spacing & 480~kHz \\
    $M$ & Number of delay grid & 64 \\
    $N$ & Number of Doppler grid & 16 \\
    $\beta$ & Rolling factor & 0.1~\cite{li_offgrid_2024}\\
    $N_t$ & Number of transmit antennas & 1024 \\
    $N_r$ & Number of receive antennas & 1024 \\
    $N_\text{RF}^t$ & Number of transmit RF chains & 4\\
    $N_\text{RF}^r$ & Number of receive RF chains & 4\\
    $N_s$ & Number of data streams & 4\\
    \bottomrule
    \end{tabular}
\end{table}

In target discovery mode, the transceiver performs beam scanning to estimate unknown sensing parameters of the target. Since the central angle of optimized sensing combiner is the assumed angle, with the knowledge of scanning angle, a new sensing combiner can be regenerated accordingly.

\subsection{Precoder and Communication Combiner Design}
The optimization problem in (\ref{equ49}) can generally be solved by the SVD of the channel. The precoder and communication combiner for the optimal communication performance are constructed from the first $N_s$ columns of the right and the left singular value matrices. 

Given the fact that in most cases, the sensing target is exactly the user equipment or happens to be a scatterer of communication channel, one of the RF chains in the precoder steers a beam towards the target. In this case, it is possible to achieve the optimal spectral efficiency while maintaining high accuracy sensing performance.

\section{Performance Evaluation}\label{sec:simulation}

In this section, we evaluate the performance of the ODDM waveform, four dimension sensing parameters estimation algorithm as well as the beamforming design for sensing and communication. The key simulation parameters are listed in Table~\ref{tab:parameter}. 
\subsection{Comparison of Waveforms for THz ISAC}
\begin{figure}[t]
	\centering
	\includegraphics[width=0.35\textwidth]{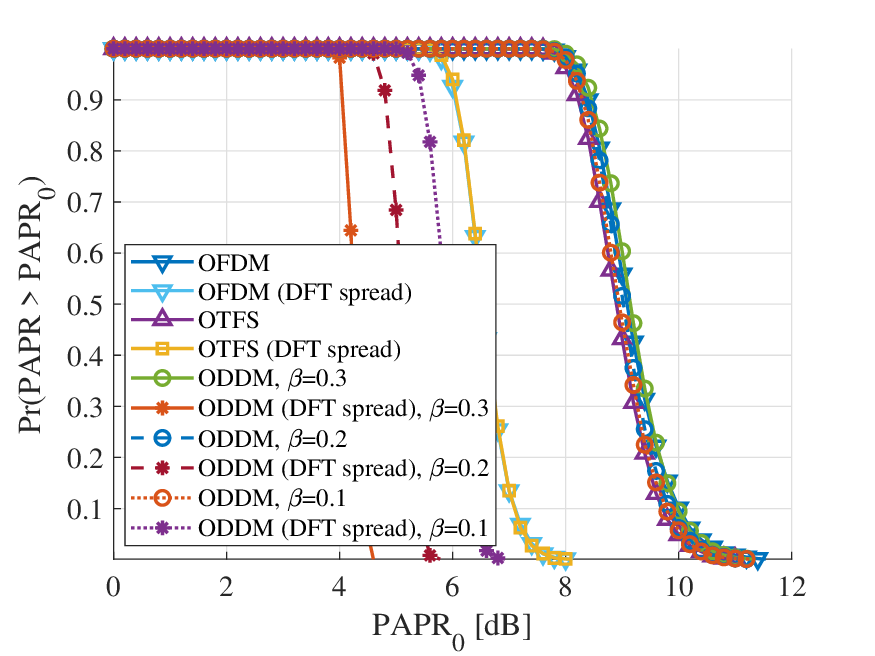}
	\caption{Comparison of peak-to-average power ratio (PAPR) of different waveforms for Terahertz sensing and communication.}
	\label{fig:wav_comp_papr}
\end{figure}

Waveforms such as OFDM, DFT spread OFDM, OTFS, and DFT spread OTFS modulate information in the time-frequency domain or the delay-Doppler domain, rather than the spatial domain. Thus, we have compared these waveforms in terms of PAPR and sensing accuracy of range and velocity, by applying a SISO system.

Fig. \ref{fig:wav_comp_papr} illustrates the PAPR of different waveforms for Terahertz sensing and communication.  The PAPR of OFDM is slightly higher than that of OTFS, while the PAPR of ODDM is nearly the same as OTFS when the rolling factor $\beta = 0.1$, and increases as the rolling factor $\beta$ increases.

The DFT spreading operation is commonly used to reduce PAPR, allowing for an approximate reduction of at least 3 dB in the PAPR for all waveforms that utilize DFT spreading. We learn that the PAPR of DFT spread ODDM is lower than that of both DFT spread OFDM and DFT spread OTFS. Numerically, the PAPR for DFT spread ODDM can be approximately reduced by 5dB ($\beta=0.1$), in contrast with ODDM. Furthermore, as the rolling factor $\beta$ increases, we observe a decrease in PAPR, which allows for flexible adjustment of the ODDM parameters based on hardware constraints.

\begin{figure}[t]
	\centering
	\subfigure[]{
		\includegraphics[width=.35\textwidth]{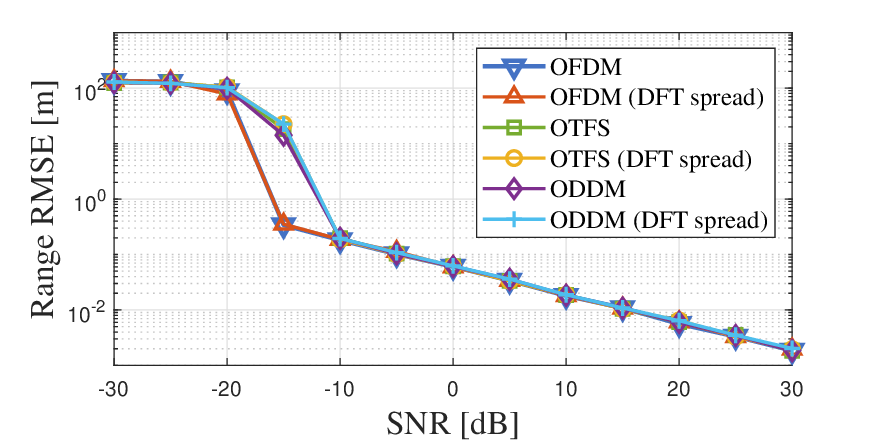}
	}
	\subfigure[]{
		\includegraphics[width=.35\textwidth]{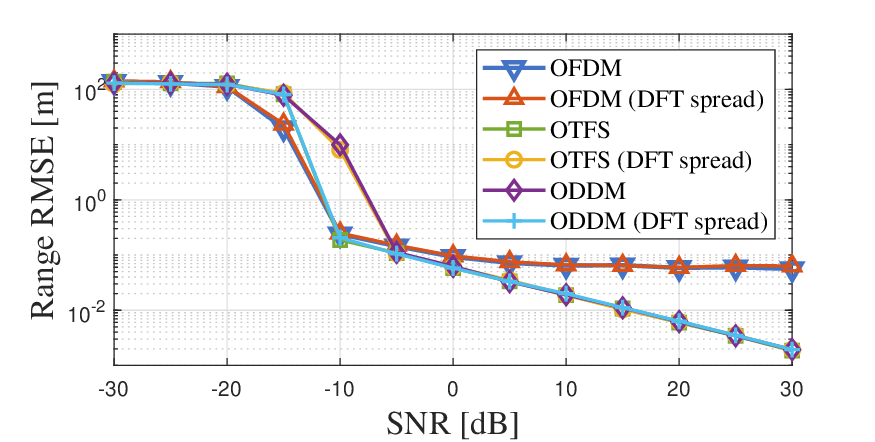}
	}	
	\caption{Root mean square error (RMSE) when the target is moving at the speed of (a) 3 km/h and (b) 300 km/h.}
	\label{fig:wav_comp_range}
\end{figure}

The high carrier frequency in the Terahertz band as well as high mobility scenarios can both induce higher Doppler effect, thus making it critical to consider the Doppler robustness of waveforms. Fig. \ref{fig:wav_comp_range} (a) represents the root mean square error (RMSE) when the target is moving at the speed of 3 km/h. We can observe that the sensing accuracies of all these waveforms are indistinguishable. However, according to Fig. \ref{fig:wav_comp_range} (b), when the target is moving at the speed of 300 km/h, the estimation accuracy of OFDM and DFT spread OFDM is degraded, while OTFS and ODDM show great Doppler robustness.

\subsection{Transmit Design: How Does Precoder Affect Sensing Accuracy?}\label{transmit_design}
\begin{figure}[t]
	\centering
	\includegraphics[width=0.43\textwidth]{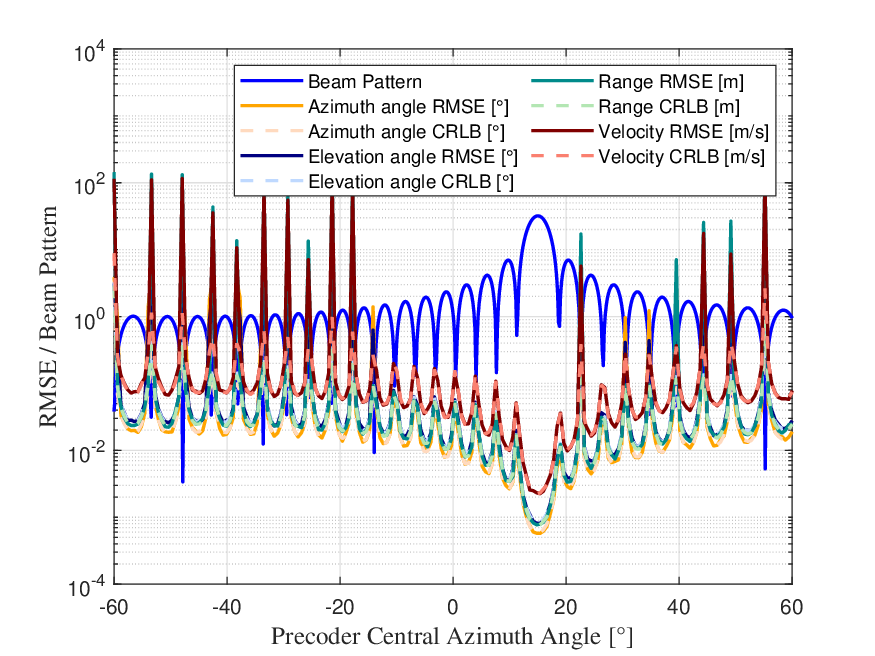}
	\caption{RMSE compared with CRLB when the central azimuth angle of precoder is different. The element of sensing combiner is randomly generated thus sensing combiner does not form a beam. The azimuth angle of target is 15$^{\circ}$. The transmit power is 20 dBm.}
	\label{fig:precoder_rmse_beampattern}
\end{figure}
Much of the literature explores the topic of beamforming for sensing and communication by designing the precoder, specifically focusing on transmit design. Thus, in this subsection, we would like to demonstrate the effect of the transmit design. For simplicity, the precoder is given by $\mathbf{F}_{s} = \frac{1}{\sqrt{N_t}} \mathbf {a}({\bar \theta},{\bar \phi}) \mathbf{1}_{N_s}^T$
where $\bar \theta$ and $\bar \phi$ are the precoder central azimuth and elevation angle. To eliminate the effect of sensing combiner $\mathbf{W} \in \mathbb{C}^{N_r \times N_s}$, each element of combiner matrix is randomly generated but satisfies $\|\mathbf{W}\|_F^2 = N_s$. In this case, this sensing combiner does not steer any beam.

Fig. \ref{fig:precoder_rmse_beampattern} illustrates the RMSE of azimuth angle, elevation angle, range and velocity with different precoders, by using the proposed estimation algorithm in Sec. \ref{estimation_algorithm}. The parameters of the estimated target are (15$^{\circ}$, 90$^{\circ}$, 50 m, 300 km/h), and the transmit power is 20 dBm. The estimation accuracy is compared with CRLB to evaluate the accuracy of the proposed sensing algorithm. We can observe that in most cases, the sensing accuracies of the four parameters are sufficiently close to their corresponding CRLBs. 

The estimation accuracy is highest when the central angle of the precoder is aligned with the true target angle. In this case, the angle estimation can reach milli-degree level and the range and velocity can reach millimeter level and millimeter-per-second level. By changing the central angle of the precoder, the transmit signal power towards corresponding direction is changed and then influences the sensing accuracy.

The blue line in Fig. \ref{fig:precoder_rmse_beampattern} gives a direct visualization of the beam pattern of the precoder, whose central azimuth angle is 15$^{\circ}$. We can observe that the RMSE for different central angles of the precoder is approximately the reverse of the beam pattern. This observation validates the claim presented in Sec. \ref{CRLBderive}, that the precoder affects the sensing accuracy by affecting the signal power in the target direction. Thus, the best sensing precoder is the one that can steer all beams towards the target to achieve the highest power, given by (\ref{equ83}).

\subsection{Receive Design: Sensing Combiner Optimization}
\begin{figure}[t]
	\centering
	\includegraphics[width=0.47\textwidth]{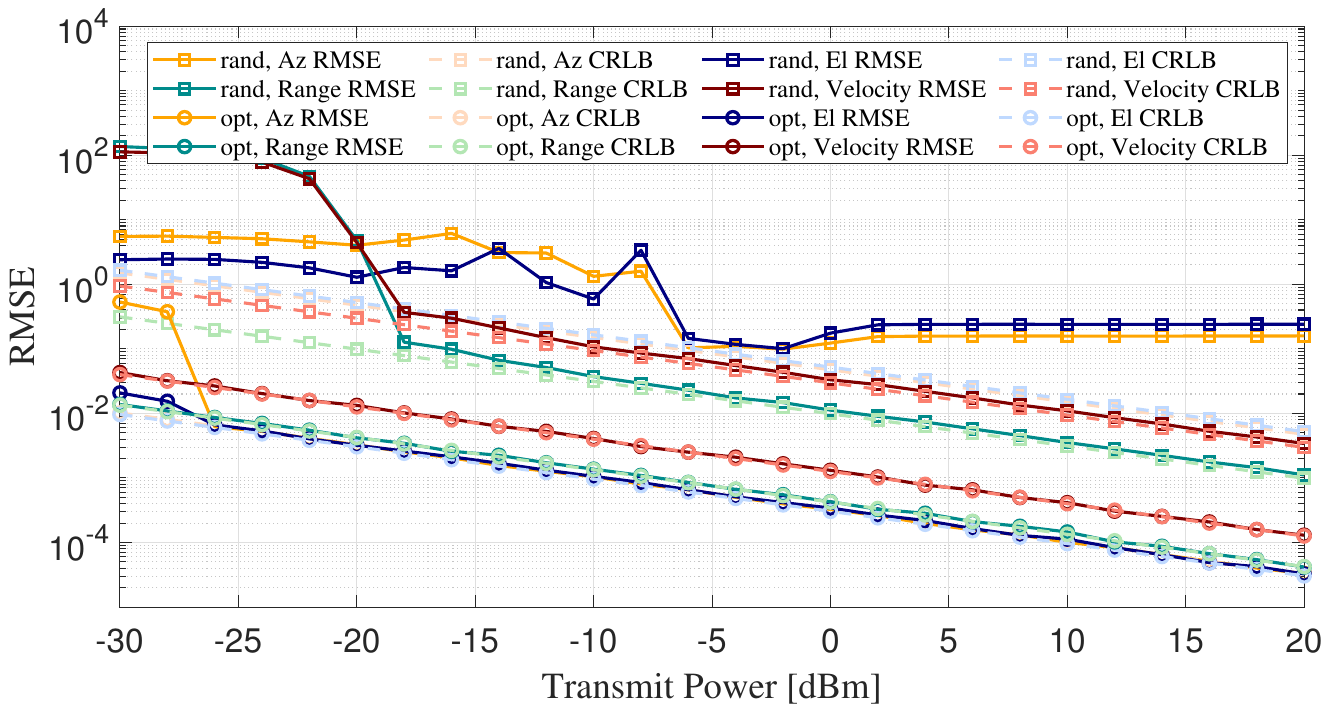}
	\caption{Azimuth (Az) angle, elevation (El) angle, range and velocity estimation accuracy with preset precoder and random or optimized sensing combiner. Note that the random case implies that the beam steering angle of $N_s$ RF chains in sensing combiner is randomly chosen.}
	\label{fig:combiner_opt_rmse}
\end{figure}
\begin{figure}[t]
	\centering
        \includegraphics[width=0.43\textwidth]{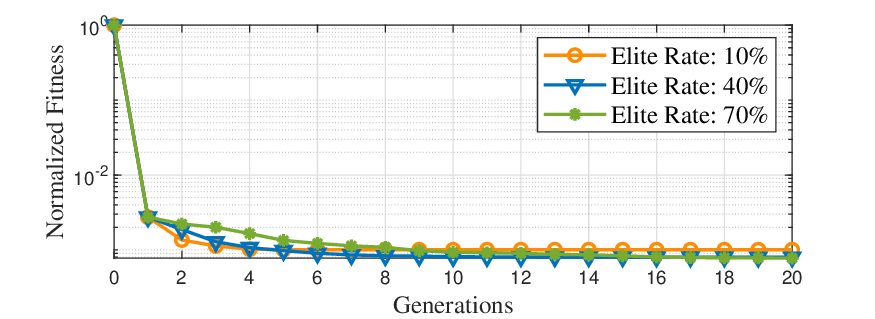}
	\caption{The convergence of the genetic-based sensing combiner optimization algorithm.}
	\label{fig:combiner_opt_algo}
\end{figure}

Fig. \ref{fig:combiner_opt_rmse} evaluates the azimuth angle, elevation angle, range and velocity estimation accuracy with a randomly chosen sensing combiner and optimized sensing combiner by applying the algorithm proposed in Sec. \ref{CRLB_opt_algo}. Note that the random case, different from the case in Sec. \ref{transmit_design} that does not steer any beam, implies that the beam steering angle of $N_s$ RF chains in sensing combiner is randomly chosen. The precoders used for two different sensing combiners are the same, given by (\ref{equ83}), considered as the best sensing precoder. According to Fig. \ref{fig:combiner_opt_rmse}, the angle estimation accuracy of a randomly chosen sensing combiner cannot approach its CRLB. The angle CRLB of the optimized sensing combiner could be two orders of magnitude lower than that of a randomly chosen sensing combiner. Similarly, the optimization algorithm can greatly reduce the RMSE and the CRLB of estimating range and velocity.

From the perspective of the proposed algorithm, it is useful to evaluate the parameters used in the algorithm to select the best parameters for faster convergence and better accuracy. Fig. \ref{fig:combiner_opt_algo} illustrates the convergence of the algorithm with different elite rates. When the elite rate is 10\%, the algorithm shows the fastest convergence performance but a poorer fitness. Both cases when elite rate is 40\% and 70\% have the same fitness but the case where elite rate is 40\% converges faster. 

\subsection{Beam Sweeping with Optimized Sensing Combiner}\label{precoder_combiner}
\begin{figure}[t]
	\centering
        \includegraphics[width=0.47\textwidth]{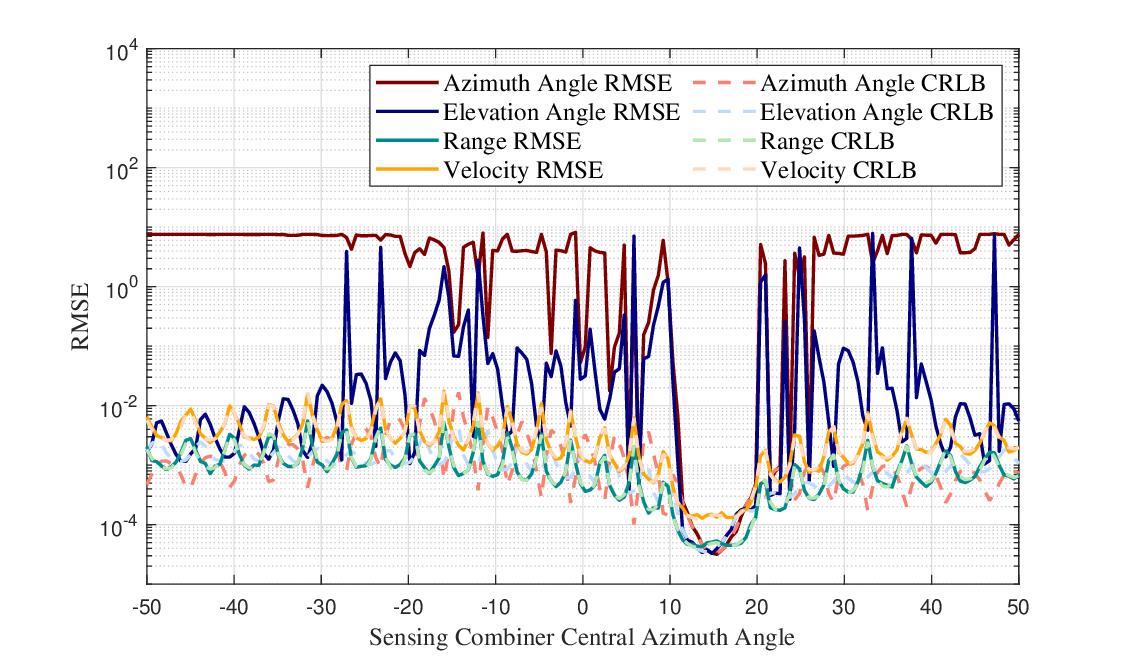}
	\caption{RMSE compared with CRLB when the precoder is fixed and the central azimuth angle of sensing combiner is different. The parameters of the estimated target is (15$^{\circ}$, 90$^{\circ}$, 50 m, 300 km/h), and the transmit power is 20 dBm. }
	\label{fig:combiner_rmse}
\end{figure}
\begin{figure}[t]
	\centering
        \includegraphics[width=0.47\textwidth]{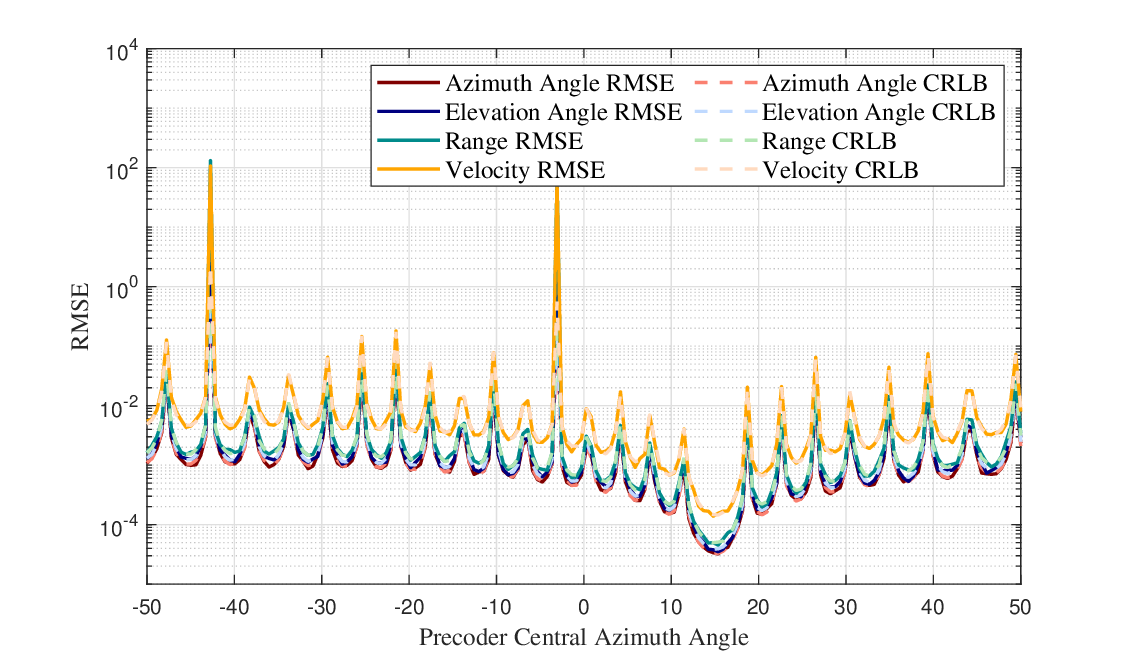}
	\caption{RMSE compared with CRLB when the sensing combiner is fixed and the central azimuth angle of precoder is different. The parameters of the estimated target is (15$^{\circ}$, 90$^{\circ}$, 50 m, 300 km/h), and the transmit power is 20 dBm. }
	\label{fig:precoder_rmse}
\end{figure}
According to Sec. \ref{CRLB_opt_algo}, the proposed optimization algorithm applies arbitrary angles to design the initial sensing combiner. As an example, the assumed azimuth and elevation angle are set as (-30$^{\circ}$, 90$^{\circ}$). 
Fig. \ref{fig:combiner_rmse} demonstrates the RMSE and CRLB of azimuth angle, elevation angle, range and velocity when the sensing combiner has different central azimuth angles and keeps on sweeping the area. Note that the sensing combiner is regenerated based on the initial sensing combiner, given the assumed angle and the central angle. In Fig. \ref{fig:combiner_rmse}, the estimation accuracy is sensitive and vulnerable with different central angles of sensing combiner. The sensing accuracy of azimuth and elevation angle can approach their CRLBs when the central angle is between 11$^{\circ}$ to 19$^{\circ}$. It means that the interval of beam scanning should be less than 8$^{\circ}$. The reason for the huge estimation error in other angle regions is that the sensing combiner can not only affect the received power but also affect the noise after the combining procedure.

With the best optimized sensing combiner from the beam sweeping procedure in Fig. \ref{fig:combiner_rmse}, we can now evaluate the sensing accuracy with precoders of different central azimuth angles, shown in Fig. \ref{fig:precoder_rmse}.
We can observe that with optimized sensing combiner, the sensing accuracy of the target can approach CRLB, except for a few outliers. Compared with Fig. \ref{fig:precoder_rmse_beampattern}, the RMSE with optimized sensing combiner is approximately one order lower than that of the sensing combiner that does not steer any beam, making the sensing accuracies with most precoder central angles satisfy the requirement of millimeter-degree level angle estimation accuracy, millimeter level range estimation accuracy and millimeter per second velocity estimation accuracy.

\subsection{Optimal Solution for Terahertz ISAC}
\begin{figure*}[t]
	\centering
	\subfigure[]{
		\includegraphics[width=.31\textwidth]{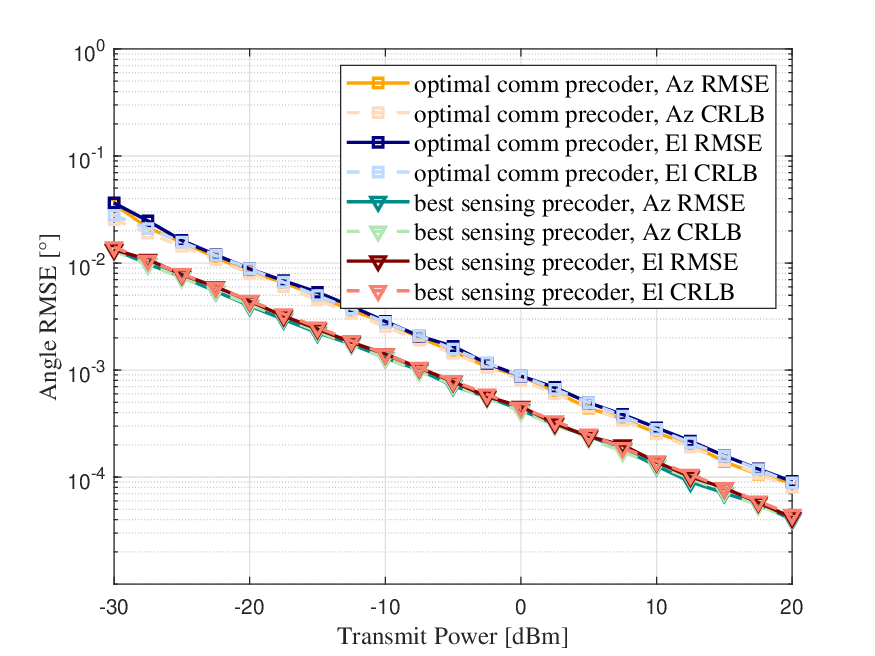}
	}
	\subfigure[]{
		\includegraphics[width=.31\textwidth]{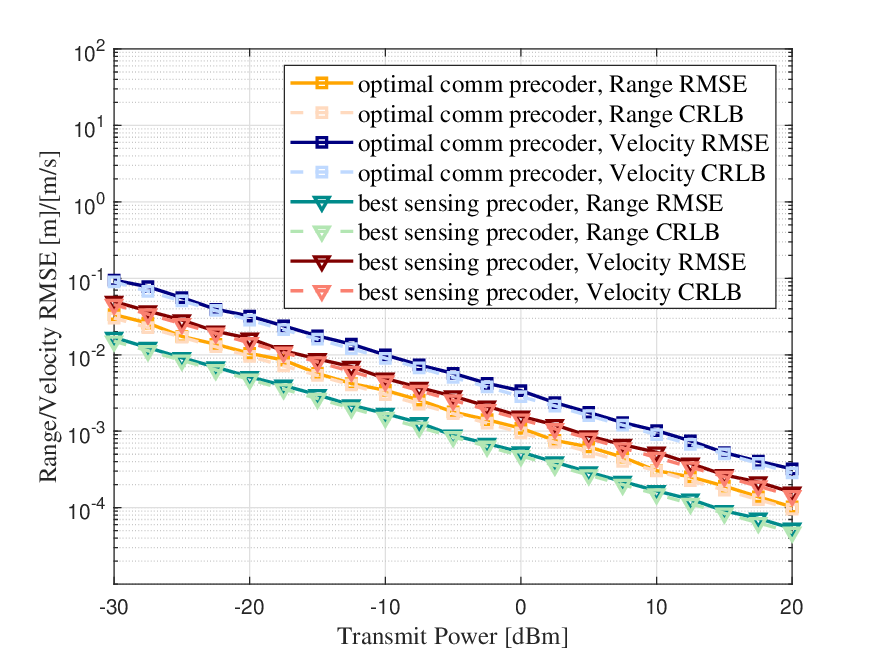}
	}
	\subfigure[]{
		\includegraphics[width=.31\textwidth]{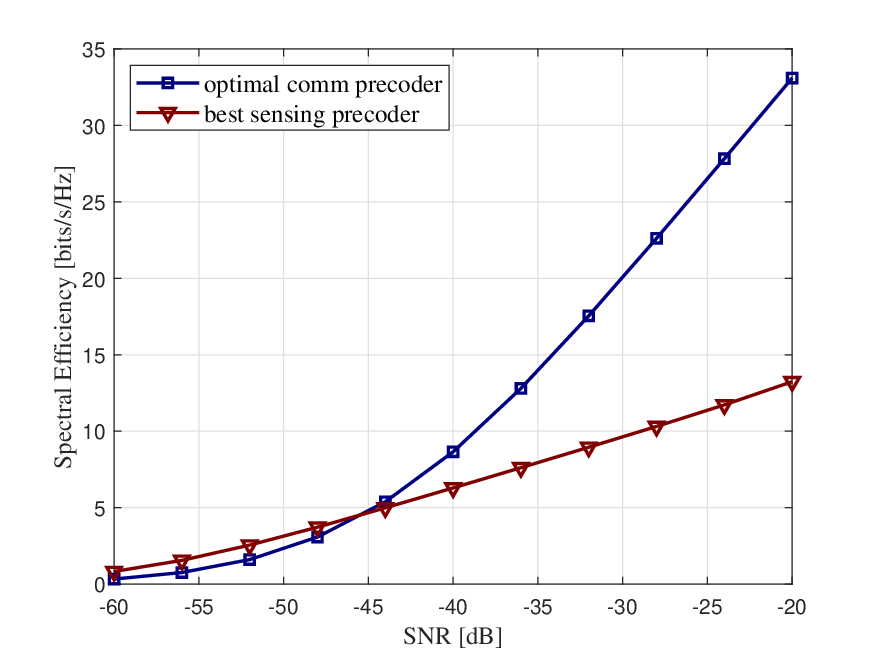}
	}	
	\caption{Evaluation of sensing and communication performance. (a) RMSE and CRLB of azimuth and elevation angle. (b) RMSE and CRLB of range and velocity. (c) Spectral efficiency. The three cases apply different precoders, but deploy same sensing combiner and same communication combiner.}
	\label{fig:radcom}
\end{figure*}
According to Fig. \ref{fig:precoder_rmse}, we have observed that with the optimized sensing combiner, the angle RMSE error of different central angles of the precoder ranges from $10^{-5}$-level degree to $10^{-2}$-level degree, except for a few outliers, which, to some extent, can be considered high sensing accuracy and is acceptable in most cases. Thus, in this subsection, we consider to achieve higher spectral efficiency while maintaining its sensing accuracy.

Fig. \ref{fig:radcom} demonstrates the sensing performance, which is evaluated by the RMSE and CRLB of estimation, and communication performance, which is evaluated by spectral efficiency with different combinations of the precoder, sensing combiner, and comunication combiner. If we apply the best sensing precoder, given by (\ref{equ83}), the optimized sensing combiner, optimized by (\ref{equ41}), as well as the optimal communication combiner, calculated by (\ref{equ49}), the transceiver possesses the combination of precoder and sensing combiner that is best for sensing. Thus, the highest sensing accuracy can be achieved with a poor spectral efficiency. This case can be considered as the ideal sensing performance.

Then we consider the case where the precoder thereinto is substituted with an optimal communication precoder. The optimal precoder and combiner for communication, given by (\ref{equ49}), is solved by the SVD of the communication channel. Given the fact that in most cases, the sensing target is exactly the user equipment or happens to be a scatterer of communication channel, one of the RF chains usually steers at the direction of the target. Thus, the root mean square error of estimating target in the best communication case is supposed to be $\frac{1}{\sqrt{N_s}}$ larger than that of the best sensing case. Numerical results show that the error for the best communication case is two times higher than that of best sensing case. Meanwhile, the spectral efficiency for the best communication case is approximately two times higher than that of the best sensing case.

Therefore, we can conclude that if we design the precoder and communication combiner by calculating the SVD, and optimize the sensing combiner by minimizing CRLB of angles, we can achieve the best spectral efficiency for communication while maintaining millimeter-level sensing accuracy. Therefore, unlike most existing studies that introduce a trade-off factor to formulate a multi-objective optimization problem, we demonstrate that the optimization problems for communication and sensing can be decoupled and solved independently, significantly reducing computational complexity.

\section{Conclusion}\label{sec:conclusion}
In this paper, we proposed a hybrid beamforming-based UM MIMO structure with ODDM modulation for Terahertz ISAC. 
To explore the integration of MIMO and delay-Doppler waveforms, we derived an off-grid signal model for ODDM and derived a UM MIMO ODDM hybrid beamforming signal model. 
To explore the sensing performance in this system, we proposed a low-complexity multi-dimension estimation algorithm that can achieve CRLB. 
To research on hybrid beamforming design for sensing and communication, we first explored how precoder affects sensing accuracy and considered to design the sensing combiner to achieve higher sensing accuracy. The results showed that by optimizing sensing combiner through the minimization of CRLB, and optimizing precoder and communication combiner through maximization of spectral efficiency, we could achieve the maximal spectral efficiency for communication while maintaining millimeter-level sensing accuracy. In light of this, the optimization problems for communication and sensing can be solved independently, which greatly reduces the computational complexity of THz ISAC system.

\appendices
\section{The derivation of FIM element}
By substitution of (\ref{equ56}) and (\ref{equ53}), the FIM element ${\mathbf J}_{\theta {\theta}}$ can be calculated with (\ref{equ45}), given by
\begin{align*}  
		{\mathbf J}_{{ {\theta {\theta}}}} 
		& = \frac{2 |\alpha |^{2}}{\sigma^2} \text{Re} \left\lbrace \text{tr} \left( 
		\text{vec}(\mathbf{W}^{H} \dot{\mathbf{A}}_{\theta} \mathbf{F} \mathbf{S})^{H} 
		\left[ \mathbf{I}_{MN} \otimes \left( \mathbf{W}^{H} \mathbf{W}\right)^{-1} \right] \right. \right.\\
		& \left. \left.\qquad
		\text{vec}(\mathbf{W}^{H} \dot{\mathbf{A}}_{\theta} \mathbf{F} \mathbf{S})
		\right)
		 \right\rbrace ,\tag{80}
\end{align*}
where $\text{vec}\left( \mathbf{W}^{H} \dot{\mathbf{A}}_{\theta} \mathbf{F} \mathbf{S}\right)$ can be further obtained as 
\begin{equation}
\setcounter{equation}{81}
	\text{vec}\left( \mathbf{W}^{H} \dot{\mathbf{A}}_{\theta} \mathbf{F} \mathbf{S}\right) = 
	\left( \mathbf{I}_{MN} \otimes \mathbf{W}^{H} \dot{\mathbf{A}}_{\theta} \mathbf{F}\right) \text{vec}(\mathbf{S}),
\end{equation}
Thus, we have
\begin{align*}
	{\mathbf J}_{{ {\theta {\theta}}}} 
	& = \frac{2|\alpha |^{2} M N N_s}{\sigma^2} \text{Re} \left\lbrace 
	\text{tr} \left( \left[ \mathbf{\Delta} \mathbf{G}_{l} \mathbf{G}_{l}^{H} \mathbf{\Delta}^{*} \right] \right. \right.\\
	& \left. \left. \quad \otimes \left[ \mathbf{F}^{H} \dot{\mathbf{A}}^{H}_{\theta} \mathbf{W} \left( \mathbf{W}^{H} \mathbf{W}\right)^{-1} \mathbf{W}^{H} \dot{\mathbf{A}}_{\theta} \mathbf{F}\right] \right)
	 \right\rbrace,\tag{82}
     \label{equ87}
\end{align*}
Given that $\text{tr}\left( \mathbf{A} \otimes \mathbf{B}\right) = \text{tr}(\mathbf{A}) \text{tr}(\mathbf{B})$, (\ref{equ87}) reduces to
\begin{equation}
	\!{\mathbf J}_{{ {\theta {\theta}}}} \!= \! \frac{2 \beta |\alpha |^{2}}{\sigma^2}\!   \text{Re}\!\left\lbrace
	\text{tr}\!\left( \mathbf{F}^{H} \!\dot{\mathbf{A}}^{H}_{\theta} \mathbf{W}\! \left( \mathbf{W}^{H} \!\mathbf{W} \right)^{-1} \mathbf{W}^{H} \!\dot{\mathbf{A}}_{\theta} \mathbf{F} \!
	\right) \!
	\right\rbrace \!,\!
    \tag{83}
\end{equation}
where $\beta = M N N_s \|\mathbf{\Delta} \mathbf{G}_{l}\|_F^2$. Similarly, ${\mathbf J}_{\phi {\phi}}$ can be given by
\begin{equation}
	\!{\mathbf J}_{{ {\phi {\phi}}}} \!= \!  \frac{2 \beta |\alpha |^{2}}{\sigma^2}\!  \text{Re}\!\left\lbrace
	\text{tr}\!\left( \mathbf{F}^{H} \!\dot{\mathbf{A}}^{H}_{\phi} \mathbf{W}\!\left( \mathbf{W}^{H} \!\mathbf{W} \right)^{-1} \mathbf{W}^{H} \!\dot{\mathbf{A}}_{\phi} \mathbf{F} \!
	\right) \!
	\right\rbrace \!,\!
    \tag{84}
\end{equation}

Besides, we can derive ${\mathbf J}_{{ {\theta {\phi}}}}$ by
\begin{equation}
	\!{\mathbf J}_{{ {\phi {\phi}}}} \!= \!  \frac{2 \beta |\alpha |^{2}}{\sigma^2}\! \text{Re}\!\left\lbrace
	\text{tr}\!\left(  \mathbf{F}^{H} \!\dot{\mathbf{A}}^{H}_{\theta} \mathbf{W} \!\left( \mathbf{W}^{H} \!\mathbf{W}\right)^{-1} \mathbf{W}^{H} \!\dot{\mathbf{A}}_{\phi} \mathbf{F} \!
	\right) \!
	\right\rbrace \!,\!
    \tag{85}
\end{equation}
Another FIM element ${\mathbf J}_{{ \phi \boldsymbol{\alpha}}}$ and ${\mathbf J}_{{ \theta \boldsymbol{\alpha}}}$ can be derived as
\begin{equation*} 
	\begin{split} 
		{\mathbf J}_{{\theta \boldsymbol {\alpha}}} 
		& = \frac{2}{\sigma^2} \text{Re}\left\lbrace
		\left[ \alpha \text{vec}\left( \mathbf{W}^{H} \dot{\mathbf{A}}_{\theta} \mathbf{F} \mathbf{S}\right) \right]^{H} \right. \\
		& \quad \left. \left[ \mathbf{I}_{MN} \otimes \left( \mathbf{W}^{H} \mathbf{W} \right) \!^{-1} \right] \left[ [1,j] \otimes \text{vec}\left( \mathbf{W}^{H} \! \mathbf{A} \mathbf{F} \mathbf{S}\right) \right]
		\right\rbrace \\ 
		& = \frac{2 \beta}{\sigma^2}  \text{Re} \! \left\lbrace 
		\text{tr} \! \left(  \mathbf{F}^{H} \! \dot{\mathbf{A}}^{H}_{\theta} \mathbf{W} \left( \mathbf{W}^{H} \mathbf{W}\right)\!^{-1} \mathbf{W}^{H} \! \mathbf{A} \mathbf{F} \right) \! \alpha^{*} \! [1,j]
		\right\rbrace
	\end{split},\tag{86}
\end{equation*}
\begin{equation*}
	{\mathbf J}_{{\phi \boldsymbol {\alpha}}} = \frac{2 \beta}{\sigma^2}  \text{Re} \! \left\lbrace 
	\text{tr} \! \left(  \mathbf{F}^{H} \! \dot{\mathbf{A}}^{H}_{\phi} \mathbf{W} \left( \mathbf{W}^{H} \mathbf{W}\right)\!^{-1} \mathbf{W}^{H} \! \mathbf{A} \mathbf{F} \right) \! \alpha^{*} \! [1,j]
	\right\rbrace,\tag{87}
\end{equation*}

By substitution of (\ref{equ58}), ${\mathbf J}_{{\boldsymbol {\alpha \alpha}}}$ can be obtained as
\begin{equation*} 
	\begin{split} 
		{\mathbf J}_{{\boldsymbol {\alpha \alpha}}} 
		& = \frac{2}{\sigma^2}\text{Re}\left\lbrace
		\left[ [1,j] \otimes \text{vec}\left( \mathbf{W}^{H} \! \mathbf{A} \mathbf{F} \mathbf{S}\right) \right]^{H} \right. \\
		& \quad \left. \left[ \mathbf{I}_{MN} \otimes \left( \mathbf{W}^{H} \mathbf{W} \right) \!^{-1} \right] \left[ [1,j] \otimes \text{vec}\left( \mathbf{W}^{H} \! \mathbf{A} \mathbf{F} \mathbf{S}\right) \right]
		\right\rbrace \\ 
		& = \frac{2}{\sigma^2}\text{Re} \left\lbrace
		[1,j]^{H} \text{vec}\left( \mathbf{W}^{H} \! \mathbf{A} \mathbf{F} \mathbf{S}\right)^{H}
		\right.\\
		& \quad \left. \left[ \mathbf{I}_{MN} \otimes \left( \mathbf{W}^{H} \mathbf{W} \right) \!^{-1} \right]
		\text{vec}\left( \mathbf{W}^{H} \! \mathbf{A} \mathbf{F} \mathbf{S}\right) [1,j]
		\right\rbrace \\
		& = \frac{2}{\sigma^2}\text{Re} \left\lbrace [1,j]^{H}
		\text{tr}\left( 
		\text{vec}\left( \mathbf{W}^{H} \! \mathbf{A} \mathbf{F} \mathbf{S}\right)^{H} \right. \right.\\
		& \quad \left. \left. \left[ \mathbf{I}_{MN} \otimes \left( \mathbf{W}^{H} \mathbf{W} \right) \!^{-1} \right] \text{vec}\left( \mathbf{W}^{H} \! \mathbf{A} \mathbf{F} \mathbf{S}\right)
		\right) [1,j] \right\rbrace
		\\
		& = \frac{2 \beta}{\sigma^2} \text{Re} \left\lbrace \text{tr} \left(  \mathbf{F}^{H} \mathbf{A}^{H} \mathbf{W} \left( \mathbf{W}^{H} \mathbf{W}\right)^{-1} \mathbf{W}^{H} \mathbf{A} \mathbf{F} \right) \right\rbrace \mathbf{I}_{2}
	\end{split},\tag{88}
\end{equation*}

\ifCLASSOPTIONcaptionsoff
  \newpage
\fi



%



\bibliographystyle{IEEEtran}
\bibliography{ref_short}
%








\end{document}